\documentclass[aps,prd,twocolumn,showpacs,superscriptaddress,groupedaddress]{revtex4}  
\usepackage{graphicx}  
\usepackage{dcolumn}   
\usepackage{bm}        
\usepackage{amsmath,amssymb}   

\usepackage{epstopdf}
\usepackage{hyperref}
\hypersetup{colorlinks=true, linkcolor=blue, urlcolor=blue, citecolor=red, bookmarks=false, pdfstartview={FitH}}

\begin{document}

\widetext

\title{A simple model for explaining Galaxy Rotation Curves}

\author{Aneta Wojnar}%
 \email{aneta.wojnar@poczta.umcs.lublin.pl}
 \affiliation{\small \it
 Institute of Physics, Maria Curie-Sklodowska University, 20-031 Lublin, pl. Marii Curie-Sklodowskiej 1, Poland}
\author{Ciprian A. Sporea}%
 \email{ciprian.sporea@e-uvt.ro}
\affiliation{\small \it
 West University of Timi\c soara, V.  P\^ arvan Ave.  4, RO-300223 Timi\c soara, Romania}
\author{Andrzej Borowiec}
 \email{andrzej.borowiec@ift.uni.wroc.pl}
 \affiliation{\small \it
 Institute for Theoretical Physics, pl. M. Borna 9, 50-204, Wroclaw, Poland}

\begin{abstract}
Abstract:
  A new simple expression for the circular velocity of spiral galaxies is proposed and tested against HI Nearby
Galaxy Survey (THINGS) data set. Its accuracy is compared with the one coming from MOND.

\end{abstract}

\maketitle

\section{Introduction}

The so-called $\Lambda$CDM model, coming from slightly modified General Relativity (GR) \cite{ein1, ein2},
together with astronomical observations, indicates that there is about $30\%$ of dust matter which we know
that exists. From it we are able to detect only $20\%$ which is baryonic described by the Standard Model of particle physics. The rest of it is so-called Dark Matter
\cite{kap, oort, zwi1, zwi2, bab, rub1, rub2, cap1, cap2} which is supposed to explain the flatness of rotational galaxies' curves.
Nowadays, there are two main competing ideas for explaining the Dark Matter problem. The first one consists in modifying  the geometric part of the gravitational field equations (see e.g. \cite{cap1, nodi, nodi2}) while the other one introduces weakly interacting particles which are failed
to be detected \cite{bertone}. Despite this, it is also believed that these two ideas do not contradict each other and could be combined together in some
future successful theory. \\
If Dark Matter exists, it interacts only gravitationally with visible parts of our universe,
and it seems to also have an effect on the large scale structure of our Universe \cite{davis, refre}.
There are some models which have faced the problem of this unknown ingredient. The famous one is called Modified Newtonian
Dynamics (MOND) \cite{millg, millg2, mond1,mond2,mond3,mond4,mc1, mc2} - it has already
predicted many galactic phenomena and this is why it is very popular among astrophysicists. It has already a relativistic version:
the so-called Tensor/Vector/Scalar (TeVeS) theory of gravity \cite{bake, moffat1}. Another approach is to consider
Extended Theories of Gravity (ETGs) in which one modifies the geometric part of the field equations \cite{iorio, capzz, seba}.
There were also attempts to obtain MOND result from ETGs, see for example \cite{barvinsky, fam, bernal, barientos, brun, fares}.
The Weyl conformal gravity \cite{mannh1,mannh2,mannh3} is a next interesting proposal for explaining rotation curves. Moreover, we would also like to mention the existence of a model based on large scale renormalization group effects and a quantum effective action  \cite{rodr,rodr2, rodr3}.
In this work we will not consider any concrete theory of gravitation from which we provide the equation ruling the motion of galactic stars.
Starting from the standard form of the geodesic equation a formula for the rotational velocity will be derived. We will also present how our simple model matches the astrophysical data
and that it possesses some similarities to
ones appearing in the
literature. At the end we will draw our conclusions. The metric signature convention is $(-,+,+,+)$.

\section{Proposed model}\label{intro2}

The standard expression of the quadratic velocity for a star moving on a circular trajectory around the galactic center is simply obtained from the GR in the weak field
and small velocity approximations. One assumes that the orbit of a star in a galaxy is circular which is in a good agreement
with astronomical observations \cite{Binney}. Thus the relation between the centripetal acceleration and the velocity is simply:
\begin{equation}\label{v1}
a=-\frac{v^2}{r}.
\end{equation}
A test particle as we treat a single star in our considerations satisfies the geodesic equation
\begin{equation}\label{geod1}
\frac{d^2 x^\mu}{d s^2} + \Gamma^\mu_{\nu\sigma}\frac{d x^\nu}{d s}\frac{d x^\sigma}{d s}=0.
\end{equation}

Although the velocity of stars moving around the galactic center is very high, when compared with the speed of light, it turns out that they are still much
smaller so we deal with the condition $v<<c$. It means that in the spherical-symmetric parametrization the velocities satisfy
\begin{equation}\label{ec.3}
v^i=\left(\frac{dr}{dt}, r\frac{d\theta}{dt}, r\sin\theta \frac{d\varphi}{dt}\right)<<\frac{dx^0}{dt},
\end{equation}
where $x^0=ct$. Taking into account eq. (\ref{ec.3}) and considering the week field limit of eq. (\ref{geod1}) together with $\Gamma^0_{00}=0$ (static spacetime), we obtain
\begin{equation}\label{geod2}
\frac{d^2 x^r}{d t^2} =-c^2\Gamma^r_{0 0}.
\end{equation}
Inserting eq. (\ref{geod2}) into (\ref{v1}) one gets
\begin{equation}\label{v2}
v^2(r)=rc^2\Gamma^r_{0 0}=r\frac{d\Phi(r)}{dr}.
\end{equation}
with $\Phi(r)$ being a Newtonian potential (see for example \cite{Weinberg}) such that finally we have
\begin{equation}\label{newt}
 v^2(r)=\frac{GM}{r}
\end{equation}
where $G$ is gravitational constant while the mass $M$ is usually assumed to be $r$-dependent, that is, one deals with some
matter distribution depending on a concrete model.
Let's assume the following simple distribution of mass in a galaxy \cite{sporea}
\begin{equation}\label{massnew}
M(r)=M_0\left( \sqrt{\frac{R_0}{r_c}}\frac{r}{r+r_c} \right)^{3\beta}
\end{equation}
with $M_0$ the total galaxy mass, $r_c$ the core radius and $R_0$ the observed scale length of the galaxy. The matter distribution in eq. (\ref{massnew}) without the term containing the quare root was also used in Ref. \cite{Moffat2}.
Since the GR prediction on the shapes of galaxies curves coming from (\ref{newt}) failed against the
observation data, one looks for some modification. The first one which appears in one's mind is to consider a bit more complicated mass distribution which can also
include Dark Matter halo in his form as well as different galaxy structure, for example disk, or other shapes.

We would like to perform a bit different approach, that is, let us modify the geometry part by, for example, considering effective quantities that could be
obtained from Extended Theories of Gravity. There are many works following this approach which inspired us to examine a below toy model. The most interesting ones which
do not assume the existence of any Dark Matter according to the authors are the following:
\begin{itemize}
 \item The Modified Newtonian Dynamics (MOND) \cite{millg} (see also similar result in  \cite{mendoza} and reviews in \cite{mond1,mond2,mond3,mond4}). It is the most
 spread modification among astronomers since is very simple, does not include any exotic ingredients (Dark Matter) and the most important, it is in a
 good agreement with observations. The MOND velocity is given by
\begin{equation}\label{mm1}
v^2(r)=\frac{GM}{r}\frac{1}{\sqrt{2}}\left[1+\sqrt{1+r^4\left(\frac{2a_0}{GM} \right)^2}\, \right]^{1/2}
\end{equation}
where $a_0\approx1.2\times10^-10\,\,ms^{-2}$ is the critical acceleration. Eq. (\ref{mm1}) is obtained from the Milgrom's acceleration formula
\begin{equation}
a=\frac{MG}{r^2}\mu\left(\frac{MG}{r^2a_0}\right)
\end{equation}
using the standard interpolation function
\begin{equation}
\mu(x)=\frac{x}{\sqrt{1+x^2}}
\end{equation}
In the limit $a_{Newt}\gg a_0$, the MOND formalism gives asymptotic constant velocities
\begin{equation}\label{vc}
v^2_c=\sqrt{a_0GM}.
\end{equation}
\item Coming from $f(R)$ gravity (metric formalism) examined by \cite{capzz, capzz2}. Here, they used the ansatz $f(R)\sim R^n$, to obtain:
\begin{equation}\label{f}
v^2(r)=\frac{GM}{2r}\left[1+(1-\beta)\left(\frac{r}{r_c}\right)^\beta\right]
\end{equation}
where $\beta$ is a function of the
slope $n$ of the Lagrangian while $r_c$ is a scale length depending on gravitational system properties
\item Given by Scalar - Vector - Tensor Gravity \cite{moffat1, Moffat2} which is in very good agreement with the RC Milky Way data 
\begin{equation}\label{Moffat}
v^2(r)=\frac{GM}{r}\left[1+\alpha-\alpha(1+\mu r)e^{-\mu r}\right]
\end{equation}
where the two free parameters allow the fitting of galaxy rotation curves.
\item Our previous result \cite{sporea}, coming from Starobinsky model $f(\hat{R})=\hat{R}+\gamma\hat{R}^2$ considered in Palatini formalism which is the simplest
example of EPS interpretation
\begin{equation}\label{vstarob}
  v^2\approx \frac{GM(r)}{r}\left( 1+\frac{2GM(r)}{c^2r}-\frac{2\pi\kappa\gamma c^2r^3\rho^2}{M(r)(1+2\kappa\gamma c^2\rho)^2} \right),
\end{equation}
where we assumed the order of $\gamma$ as $10^{-10}$ taken from cosmological considerations \cite{borow}, $\rho$ is energy density obtained from mass distribution provided
by the model and (\ref{massnew}), see the details in \cite{sporea}.
\end{itemize}
We immediately observe that all these modifications coming from different models of gravity possess a feature which can be simply written as
\begin{equation}\label{result}
v^2(r)=\frac{GM}{r}\Big( 1 +A(r)\Big)
\end{equation}
where the unknown function $A(r)$ depends on the radial coordinate and some parameters. In this manner, the
function $A(r)$ is treated as a deviation from the Newtonian limit of General Relativity.

Our task now is to find a suitable function $A(r)$ which takes into account and reproduces the observed flatness of galaxy rotation curves.
Moreover, at short distances (at least the size of the Solar System)
the velocity from eq. (\ref{result}) should have as a limit the Newtonian result $v^2(r)=GM/r$. These imposes some constrains on the function $A(r)$.

\section{A particular  example}
We have seen in the previous section that there are many alternatives to General Relativity which possess extra terms that improve the behavior of the galaxy curves.
Moreover, many of them can have the same week field limit producing the same result (\ref{result}). Thus, one can explain the observed galaxy
rotation curves using the equation (\ref{result}) without the assumption on the existence of Dark Matter.

In this section we would like to propose a model for fitting the galaxy rotation curves data observed astronomically. As we will see the model fits quite well the data set of galaxies obtained from THINGS: The HI Nearby Galaxy Survey catalogue \cite{Walter,deBlok}, on which our analysis is performed.

\begin{figure*}[h!t]
\centering
\includegraphics[scale=0.293]{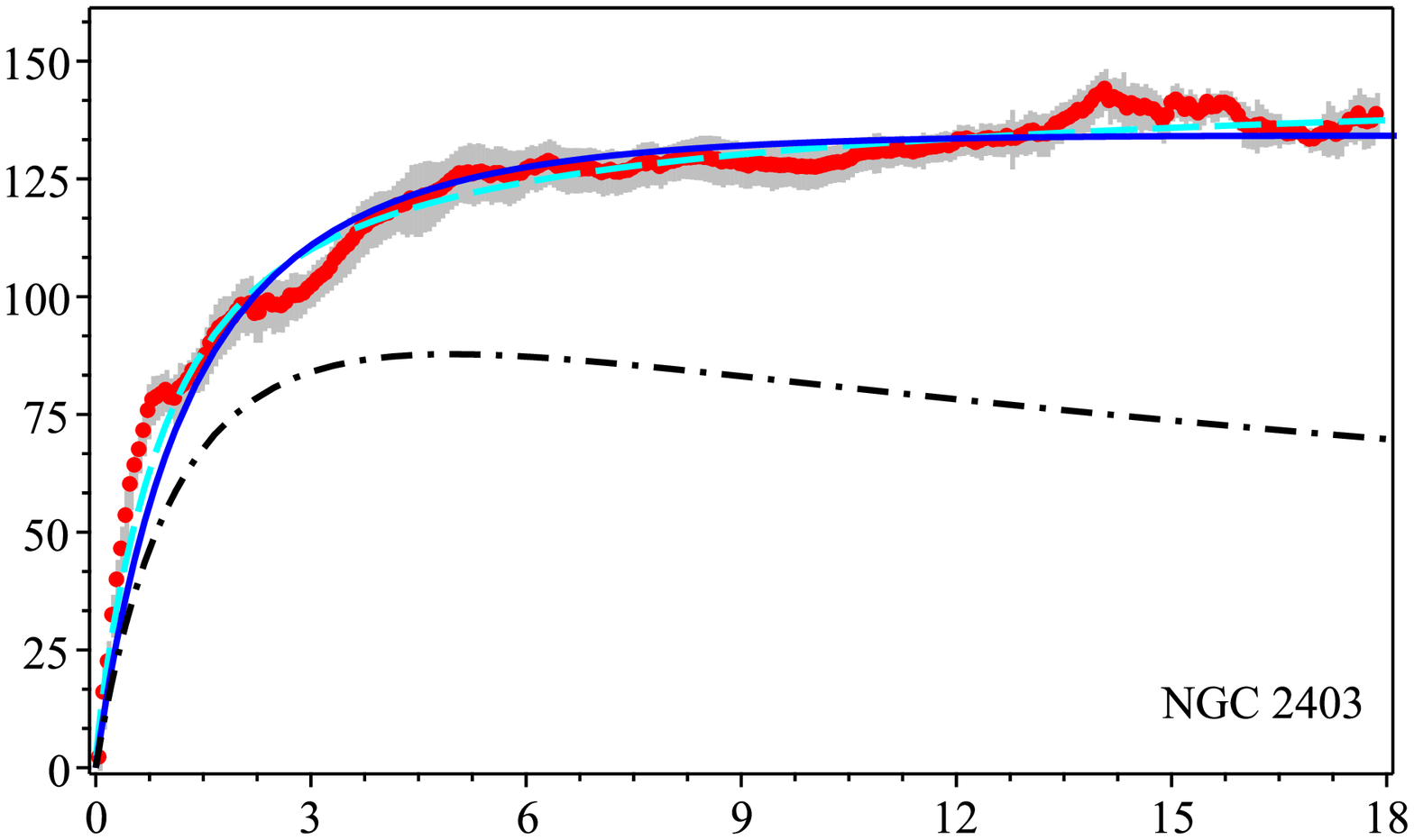}
\includegraphics[scale=0.293]{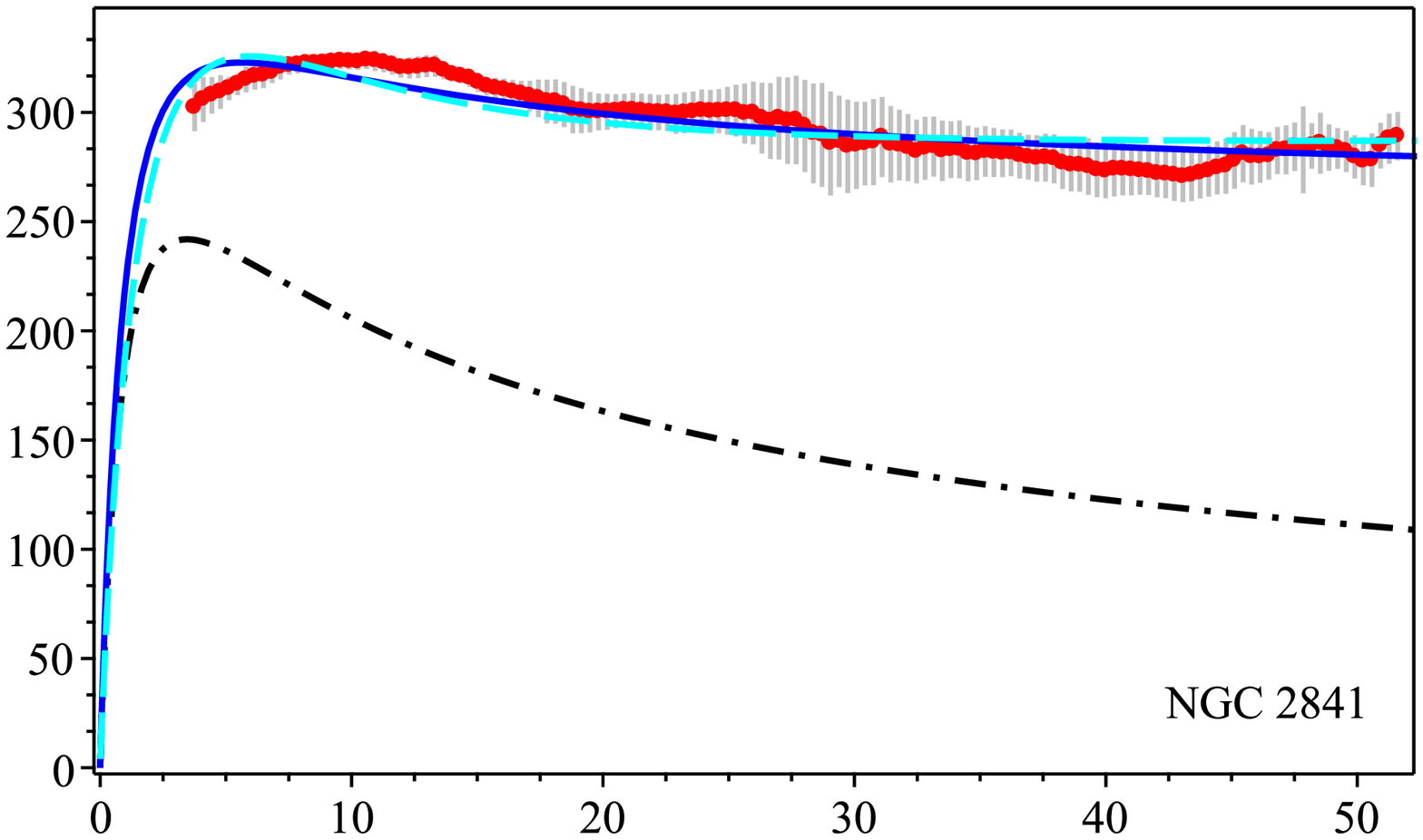}
\includegraphics[scale=0.293]{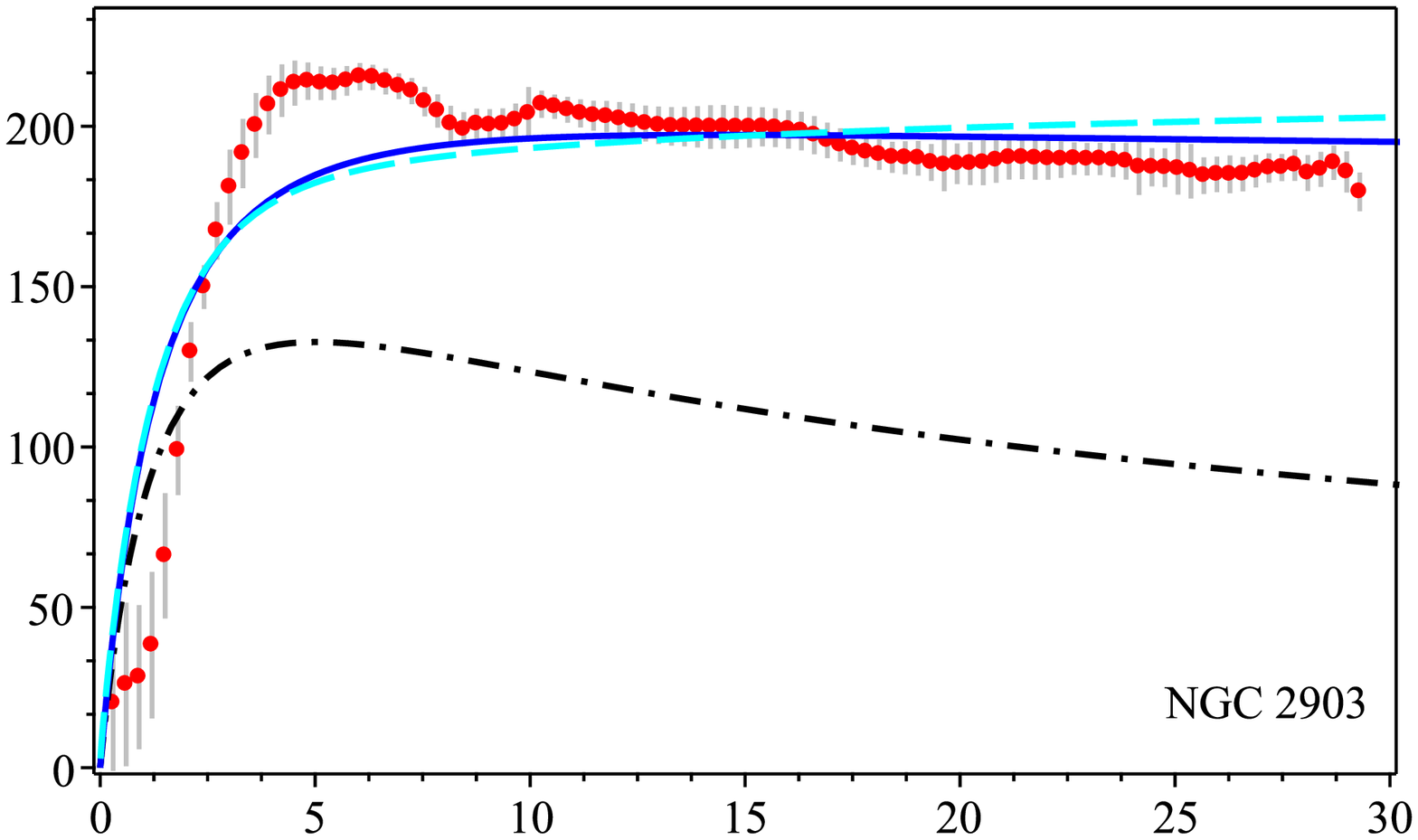}
\includegraphics[scale=0.293]{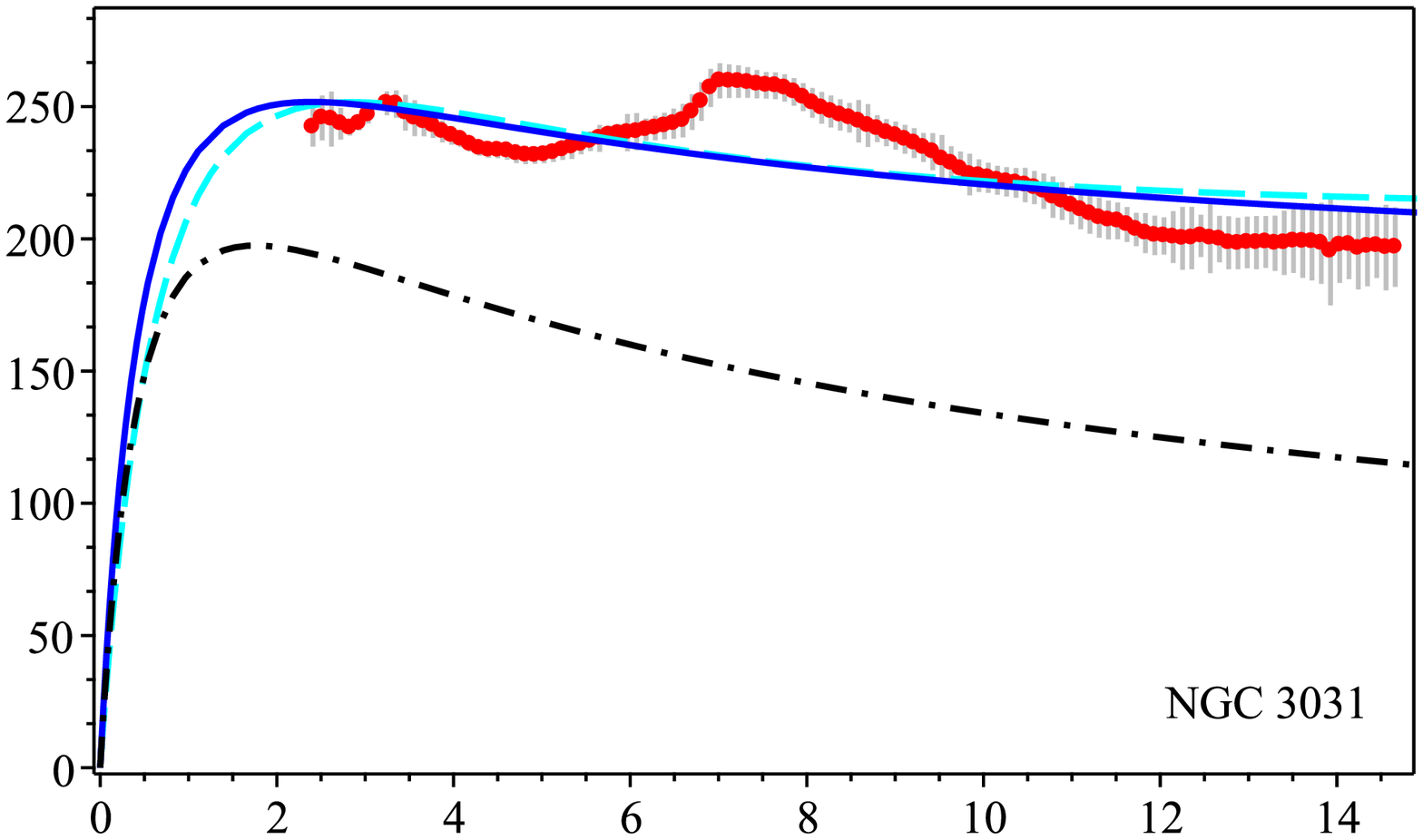}
\includegraphics[scale=0.293]{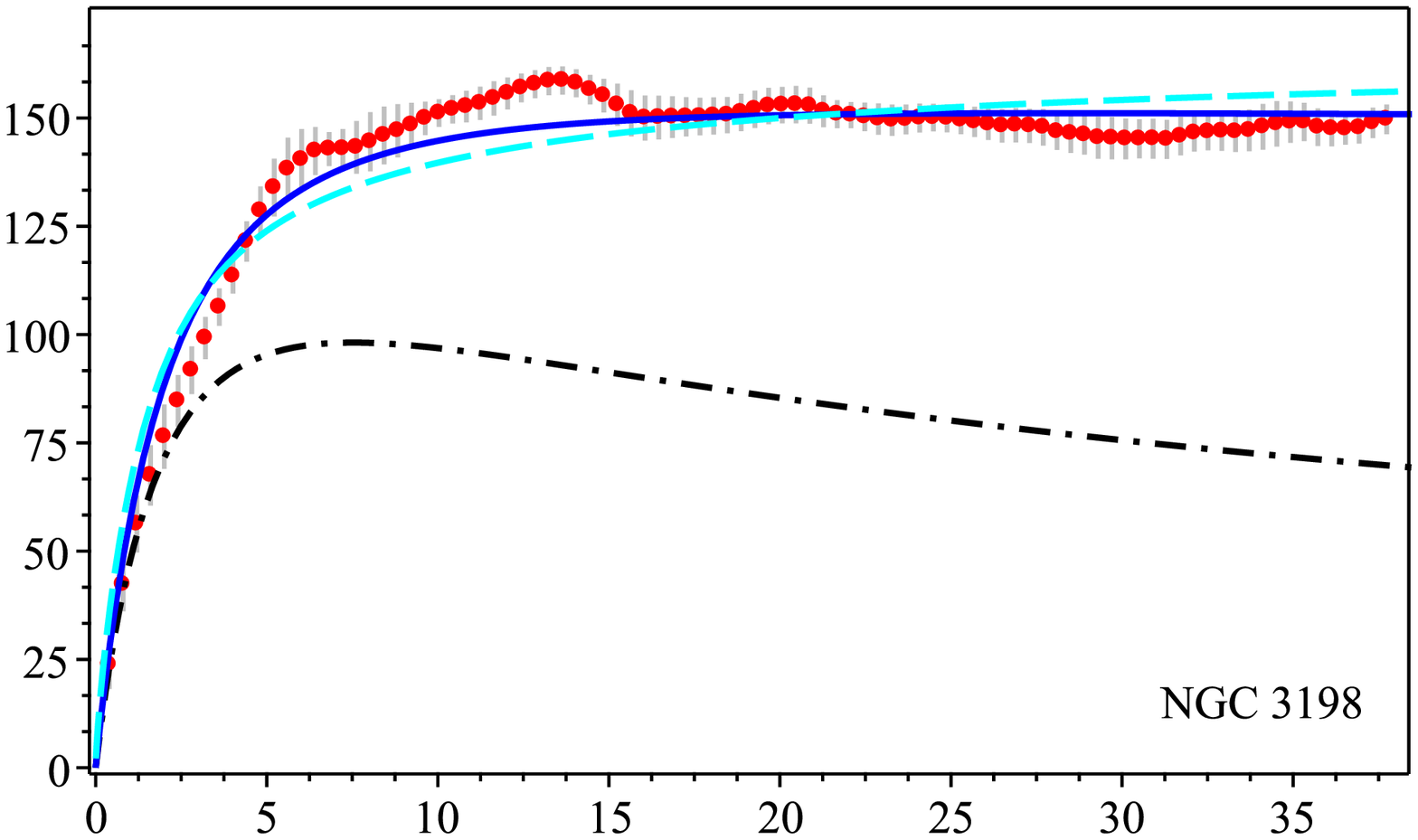}
\includegraphics[scale=0.293]{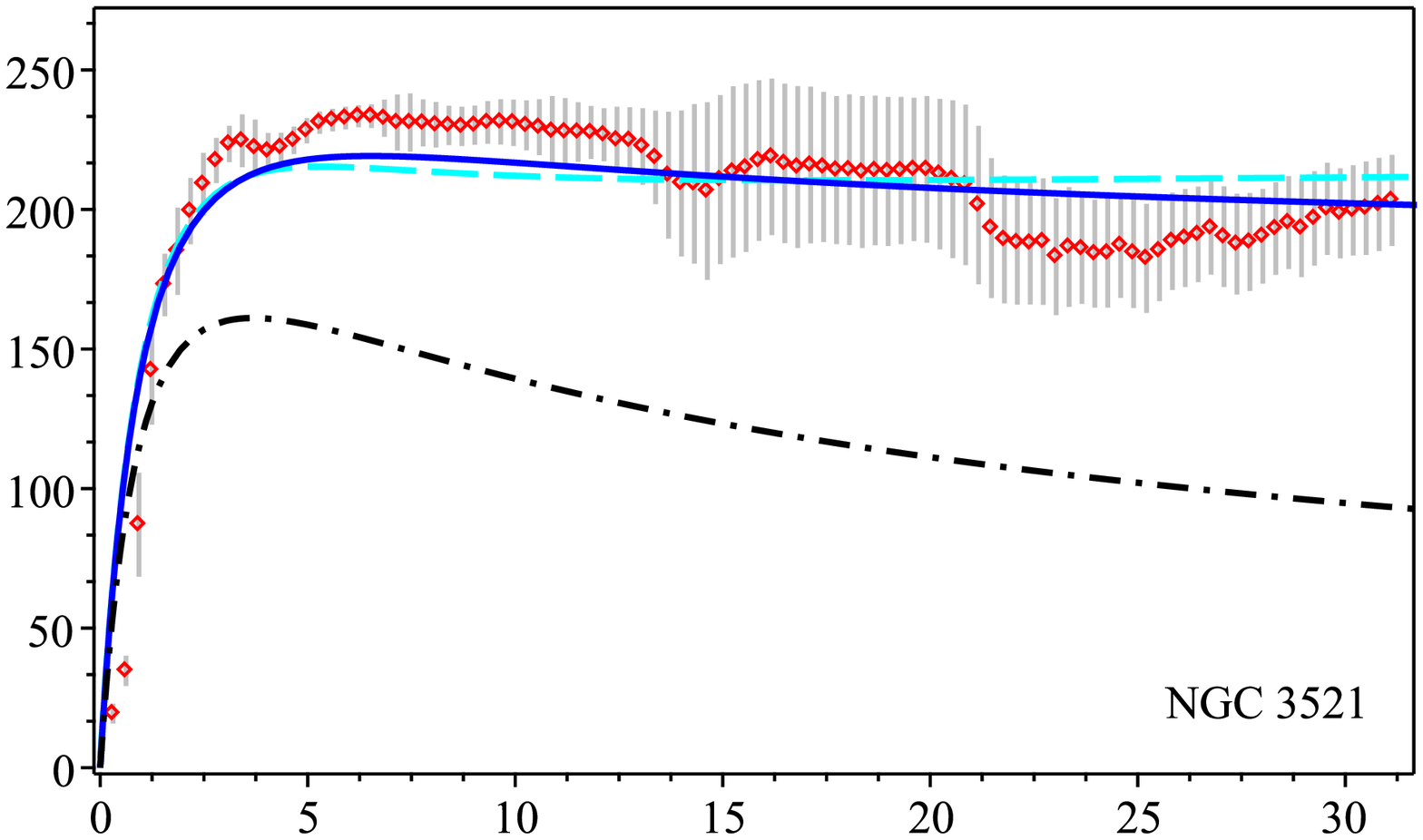}
\includegraphics[scale=0.293]{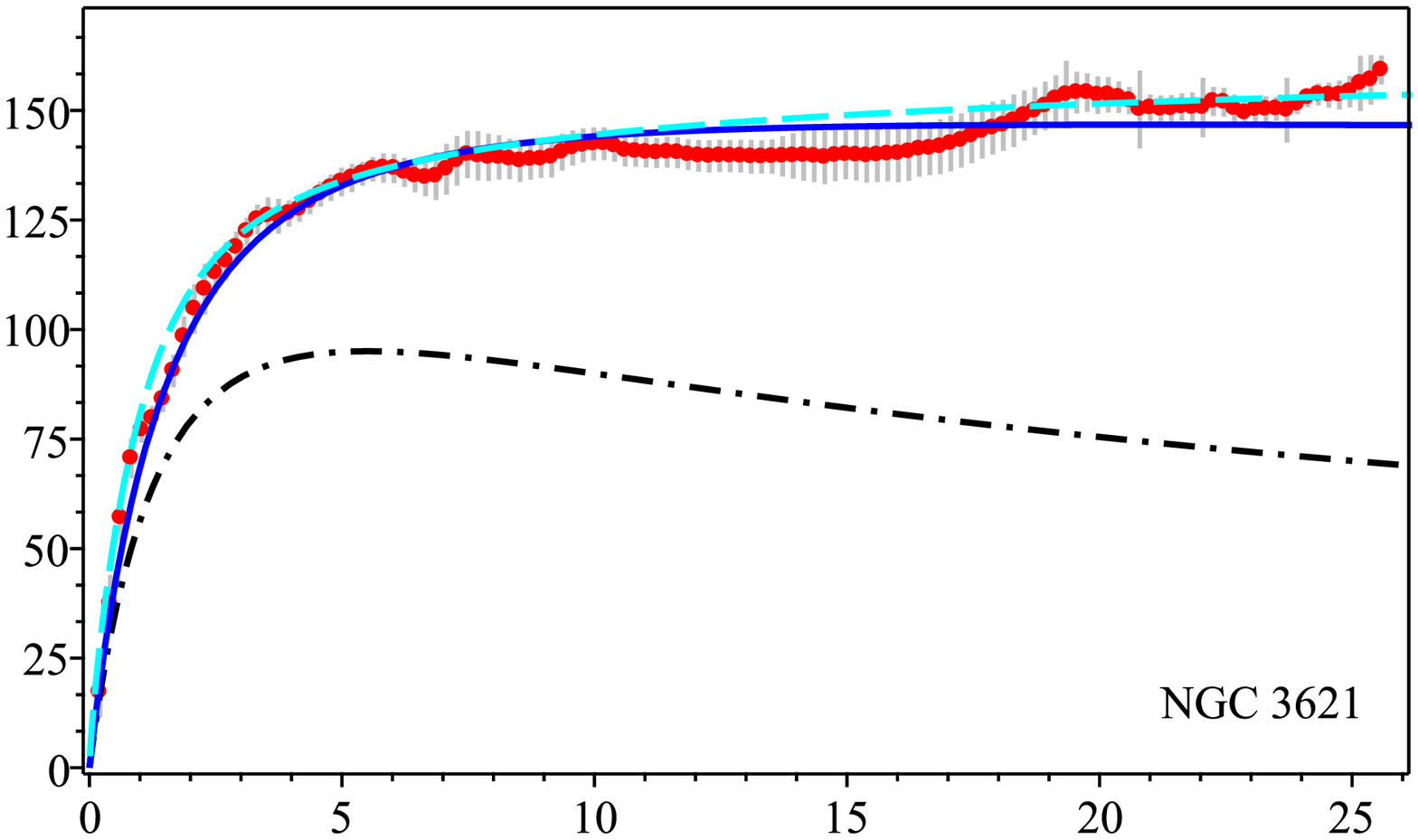}
\includegraphics[scale=0.293]{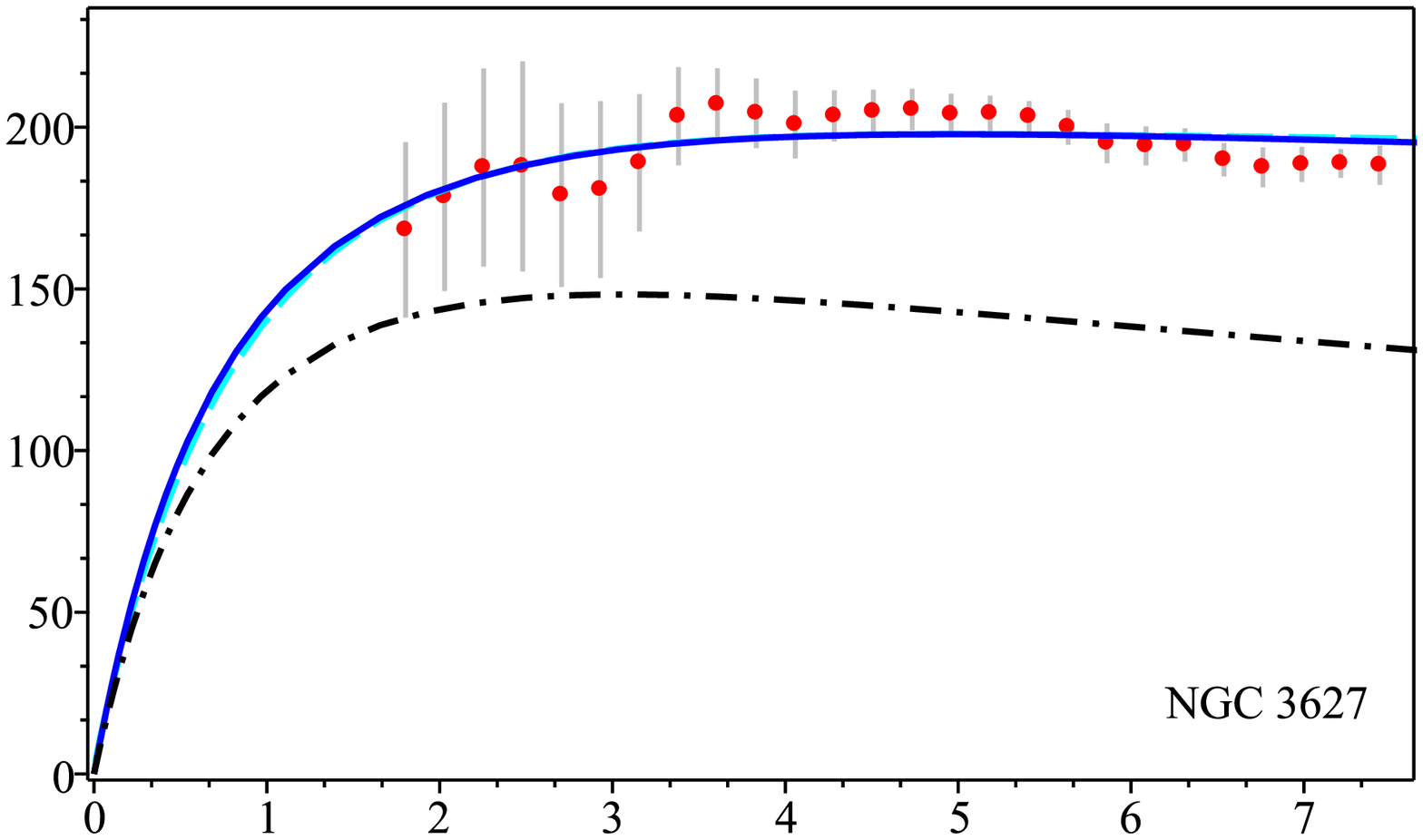}
\includegraphics[scale=0.293]{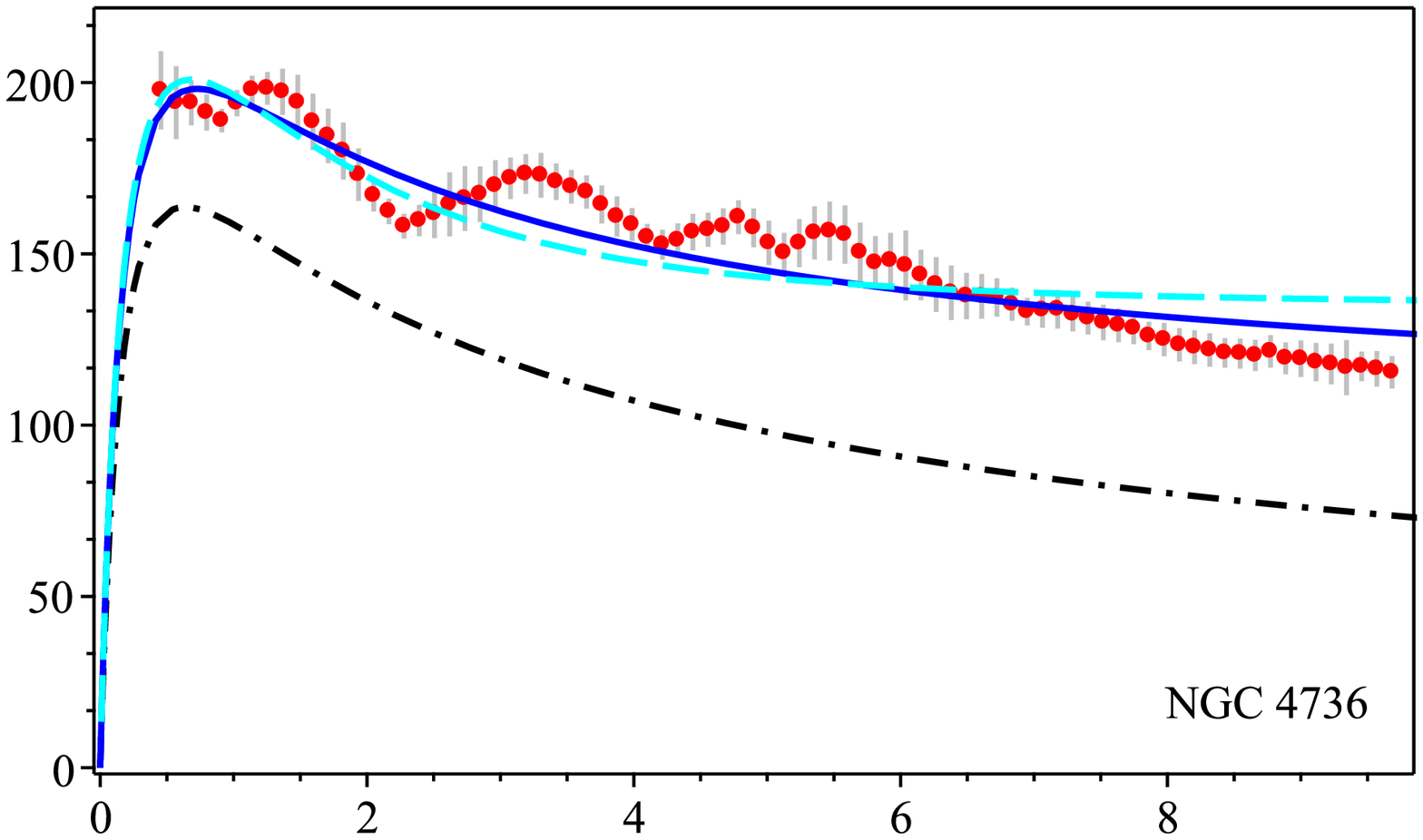}
\includegraphics[scale=0.293]{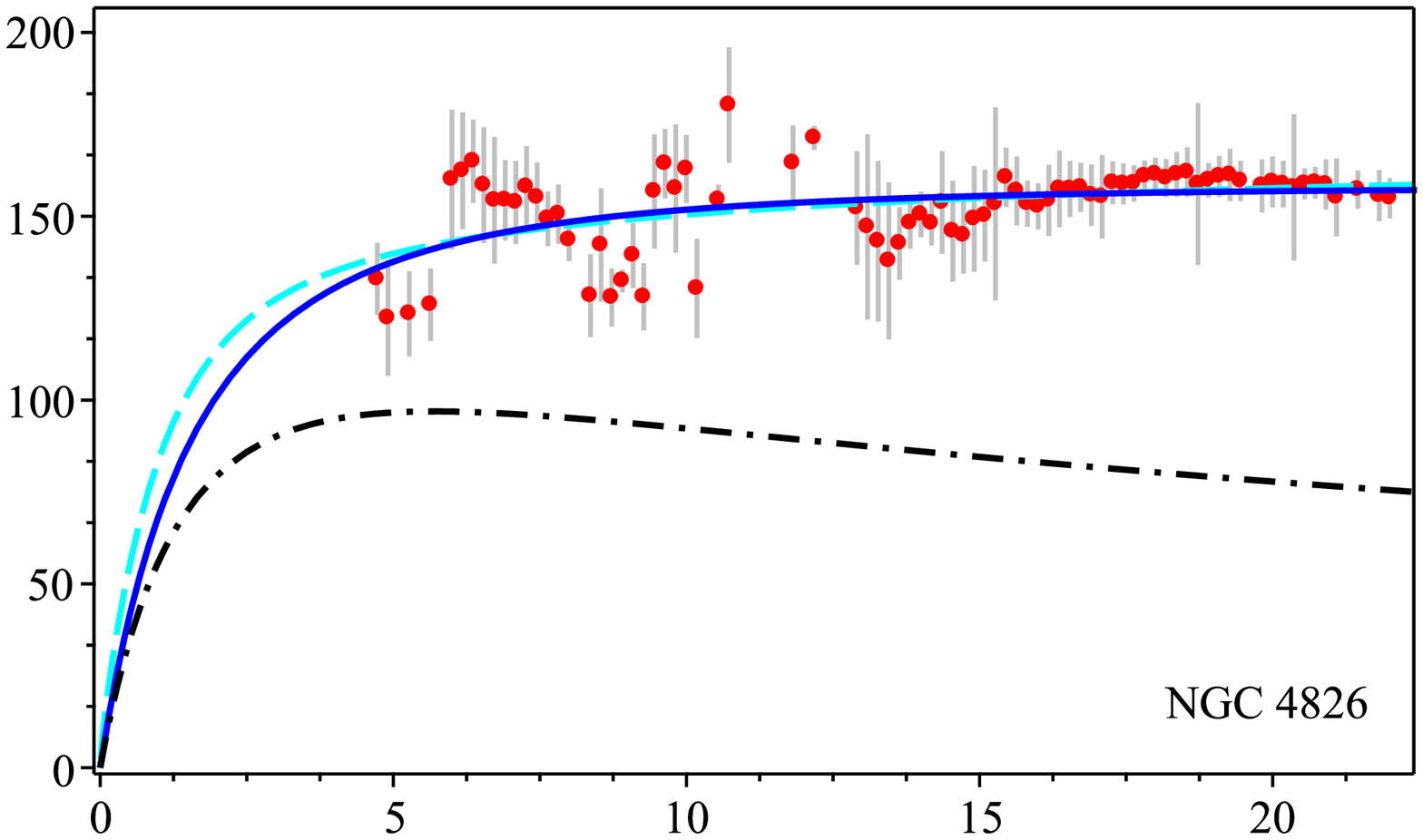}
\includegraphics[scale=0.293]{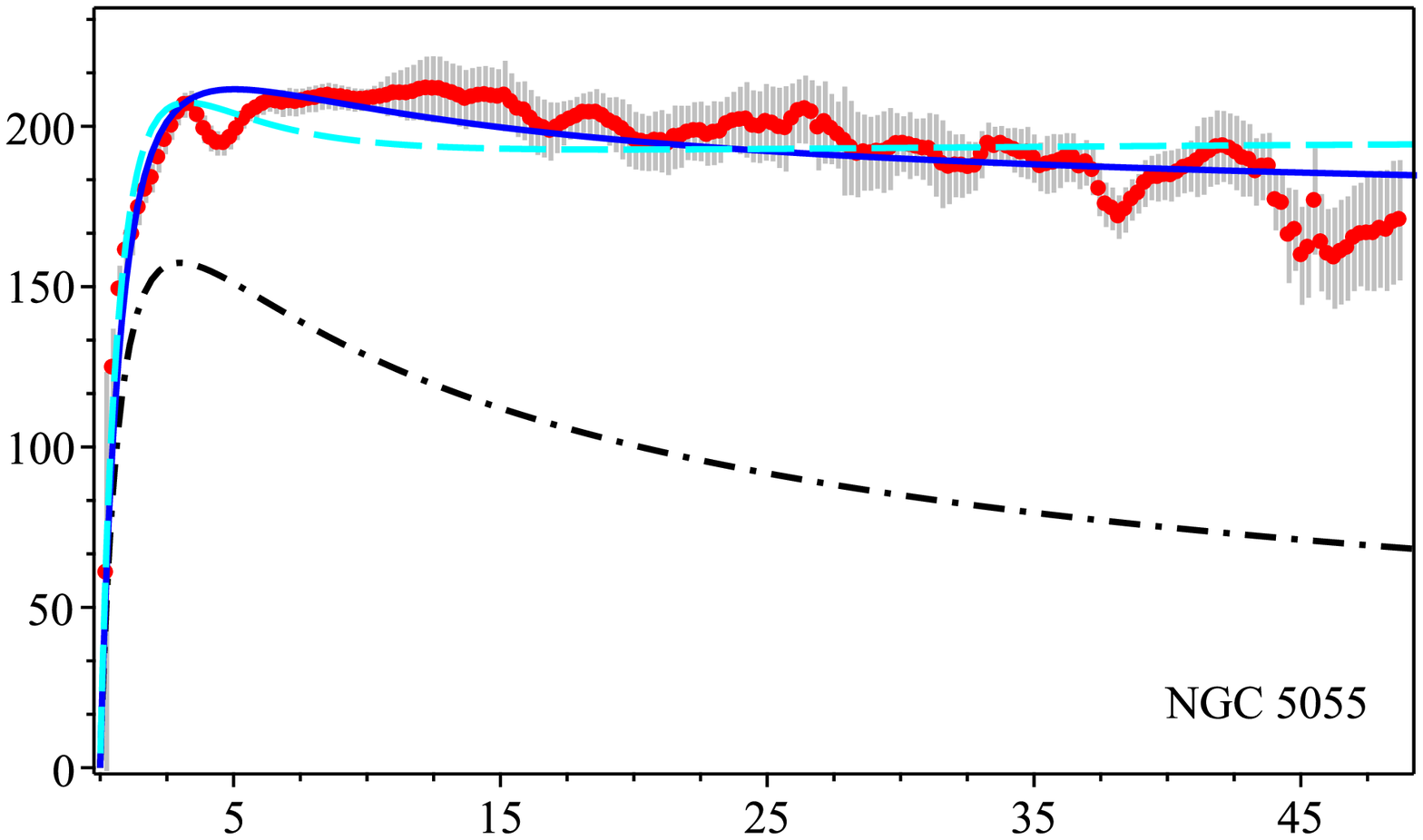}
\includegraphics[scale=0.293]{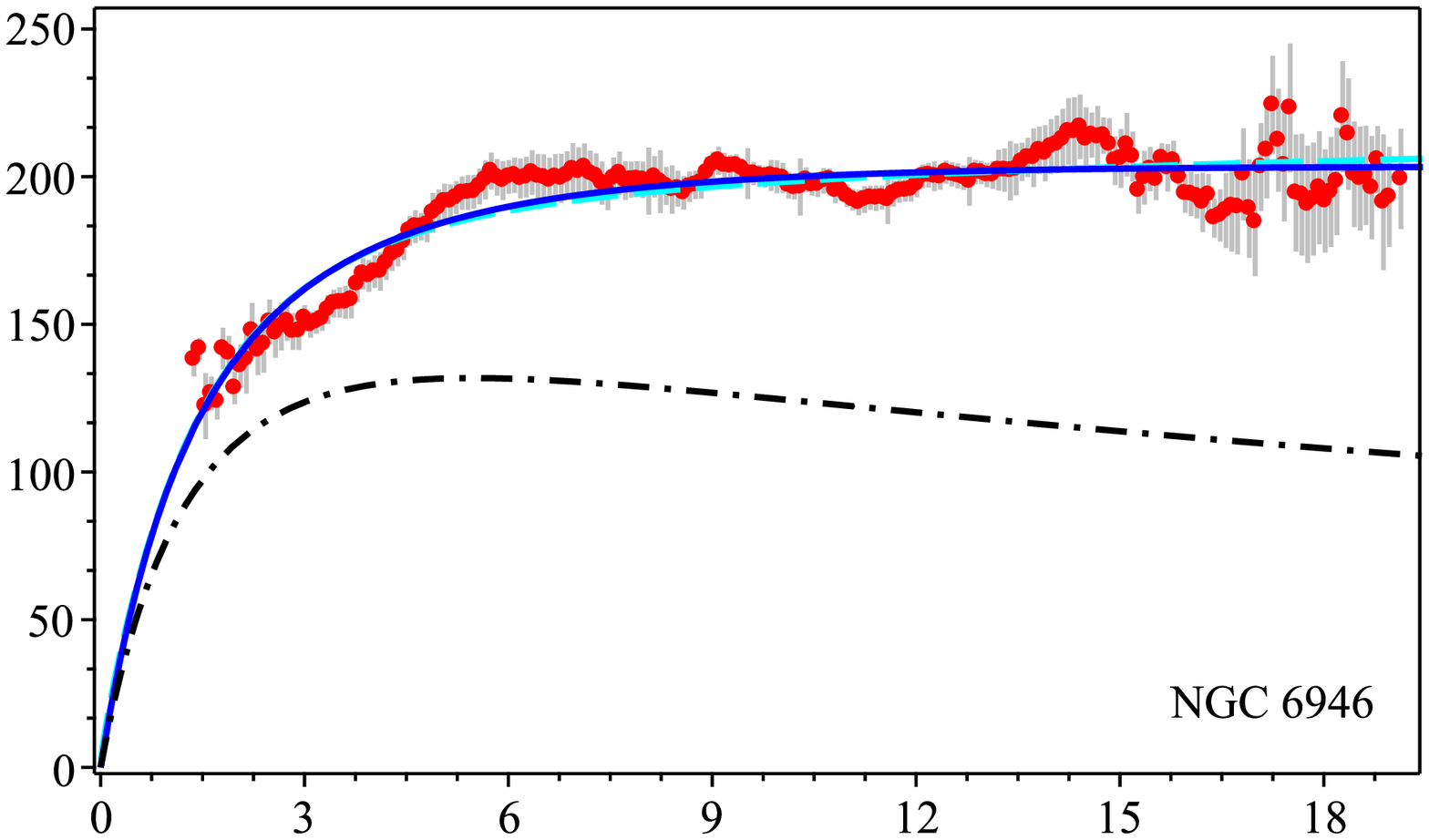}
\includegraphics[scale=0.293]{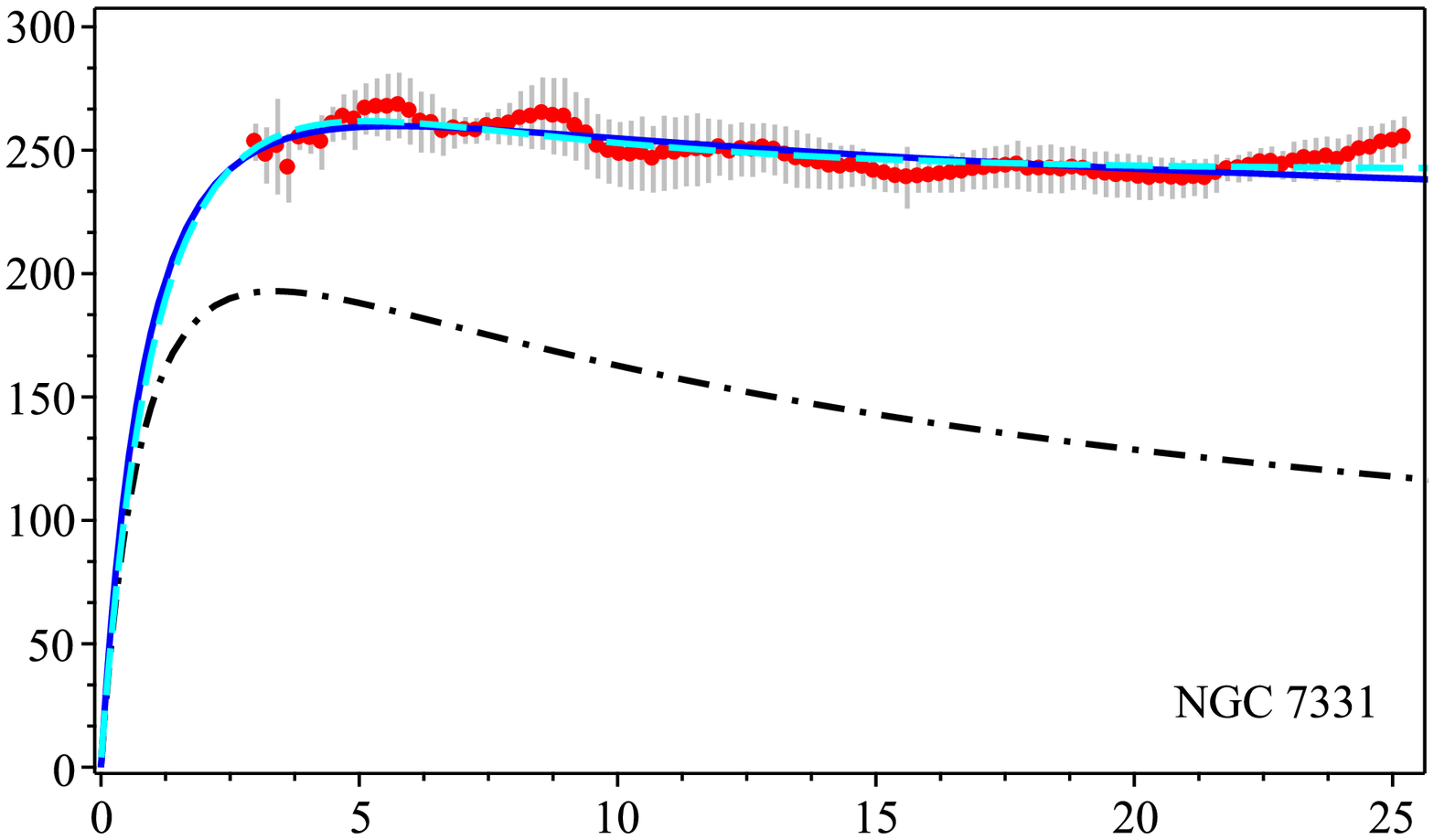}
\includegraphics[scale=0.293]{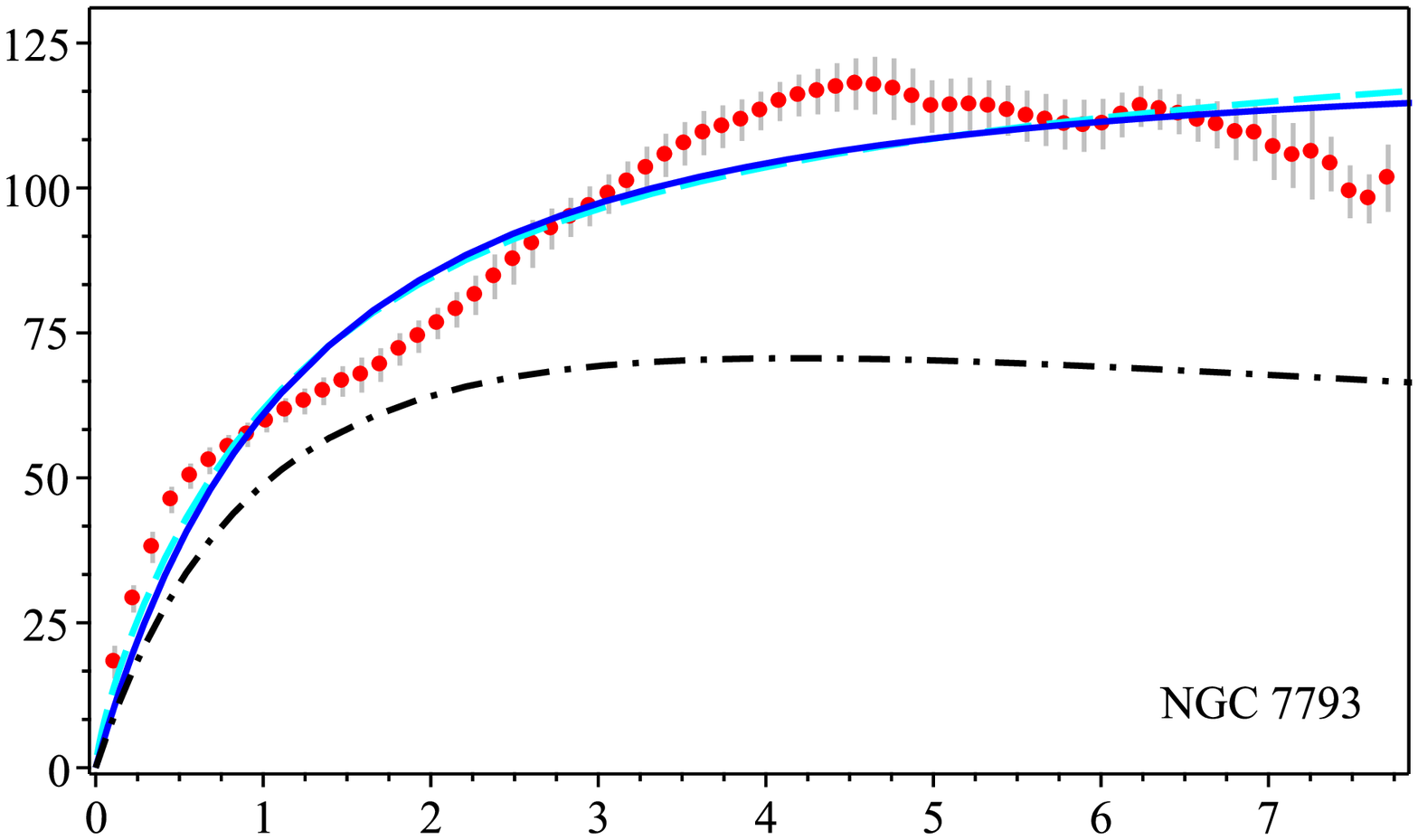}
\includegraphics[scale=0.293]{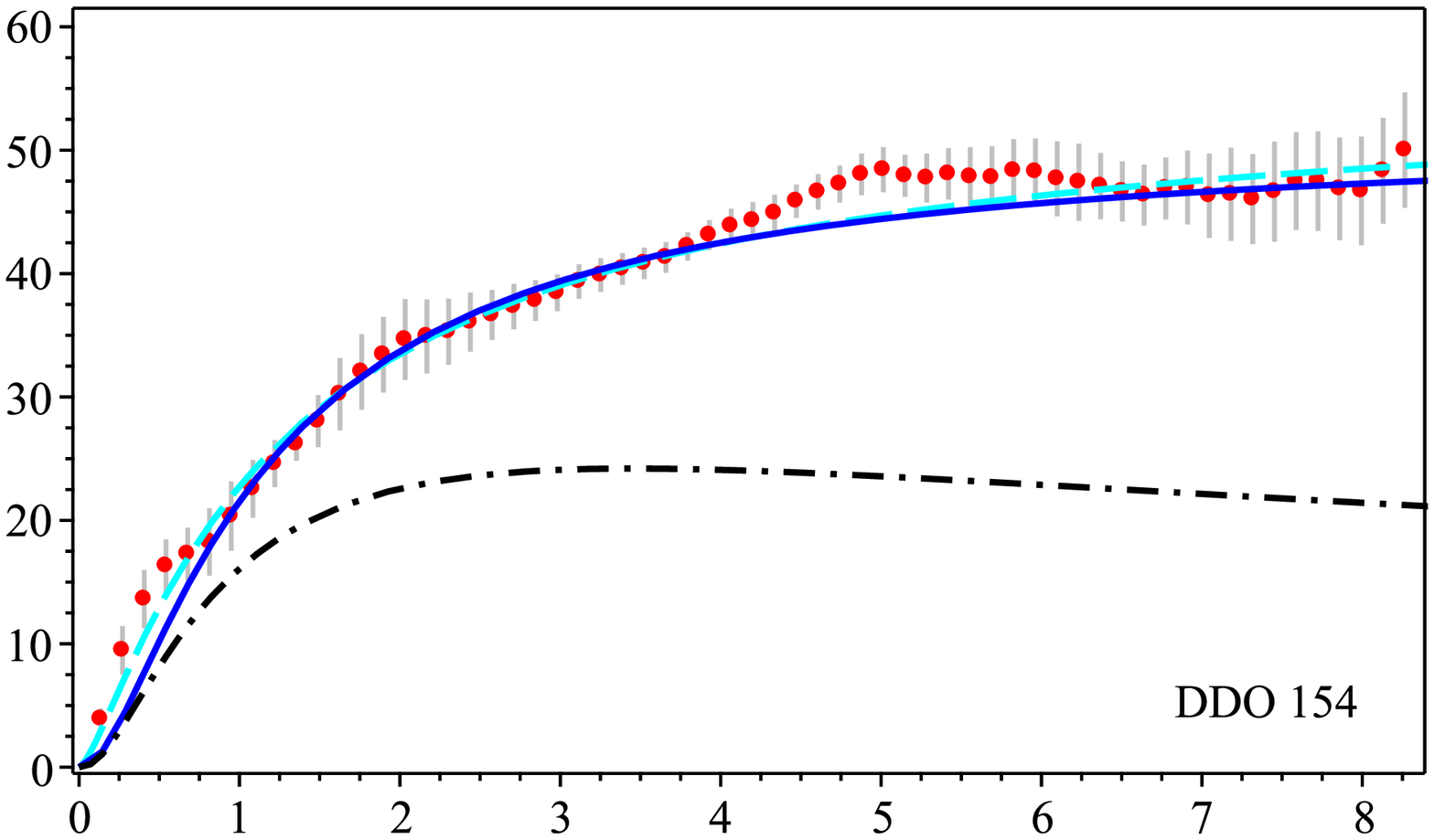}
\includegraphics[scale=0.293]{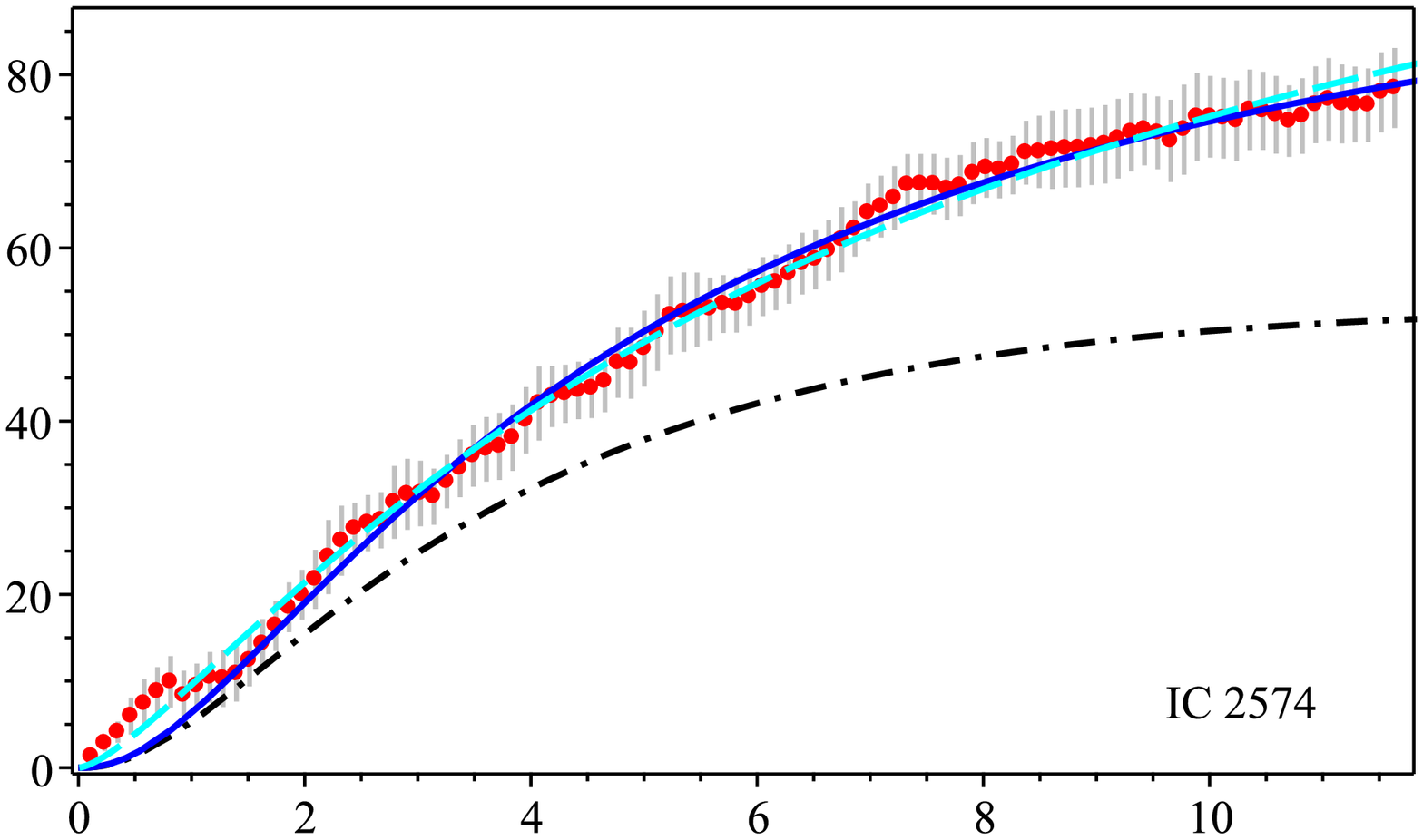}
\includegraphics[scale=0.293]{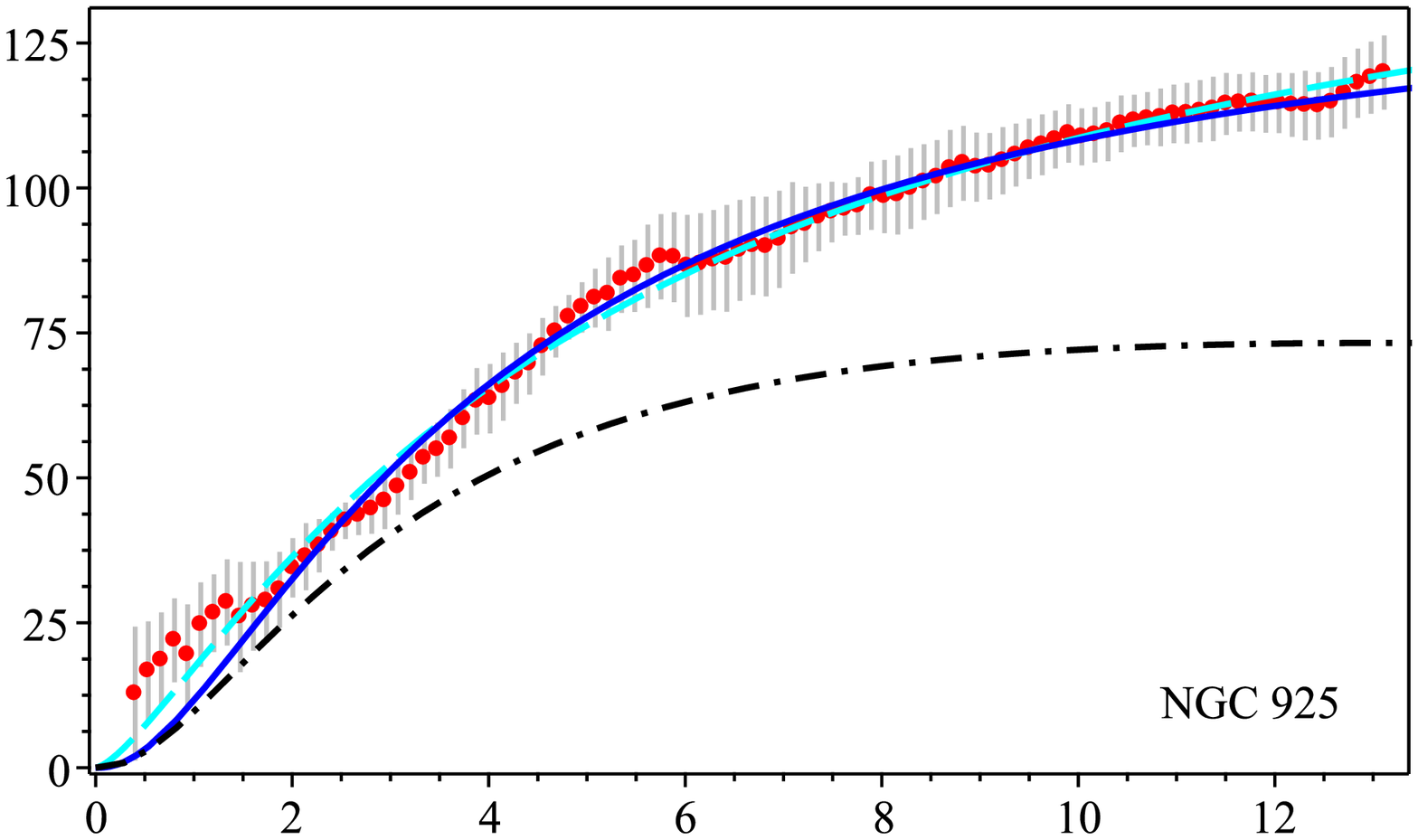}
\includegraphics[scale=0.293]{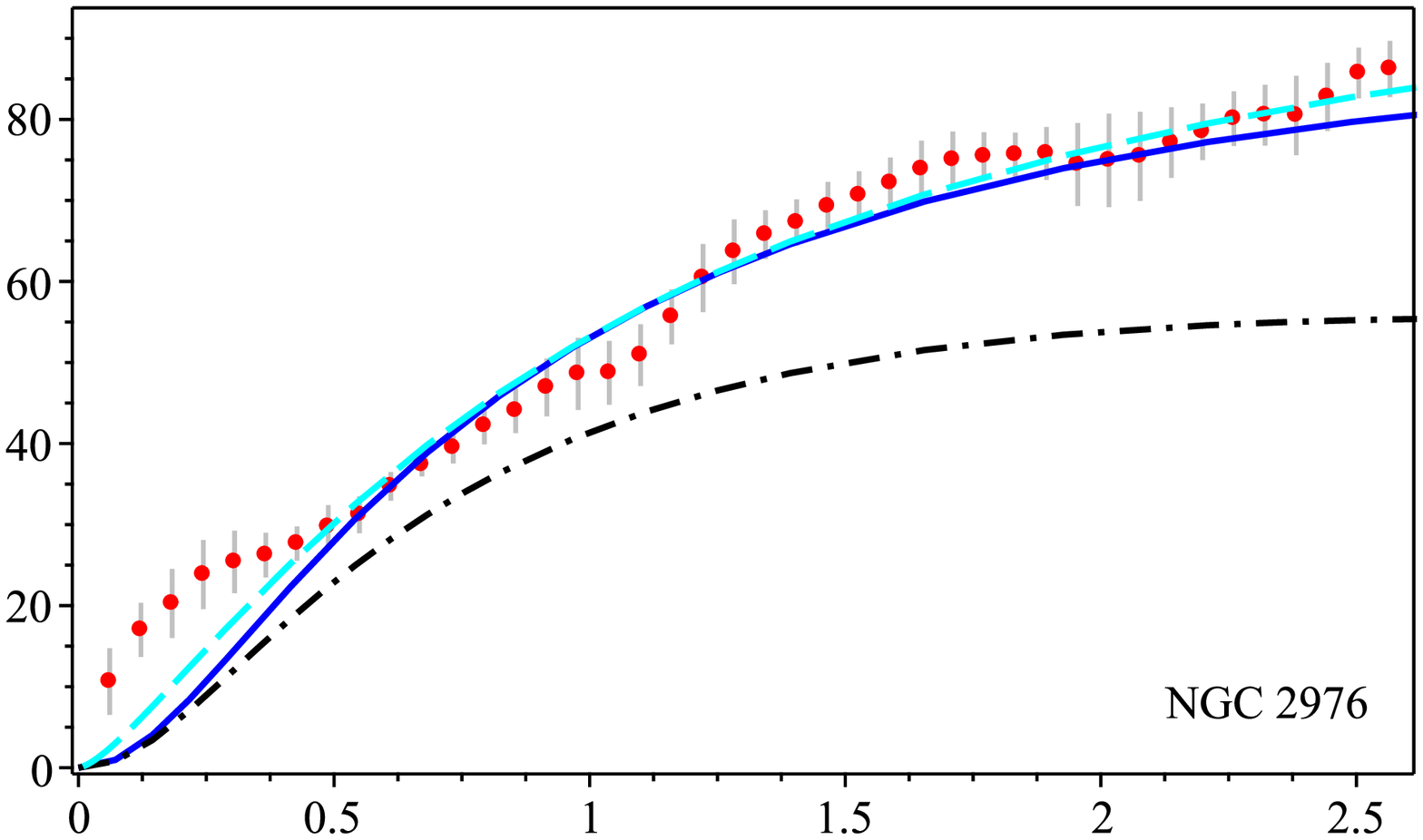}
\caption{(color online) Rotational velocities in km/s ($y$ axis) at a certain distance in kpc ($x$ axis) from the center of the galaxy. The blue curves RC are obtained from the parametric fit of eq. (\ref{vfinal}) in the case of 18 THINGS galaxies. The proprieties of the galaxies in the sample can be found in Table I from Ref.\cite{Walter}. The full (blue) curve are the rotation curves obtained using eq.(\ref{vfinal}); the (red) full circles are the observed data points where the vertical (grey) lines represent the error bars; the contribution due to the Newtonian term is given by the dash-dotted (black) lines, while the dashed (cyan) lines give the MOND rotation curves. The numerical values resulted from the fits are given in Table \ref{tab2}.}
\label{fig.2}
\end{figure*}

\begin{figure*}[h!t]
\centering
\includegraphics[scale=0.293]{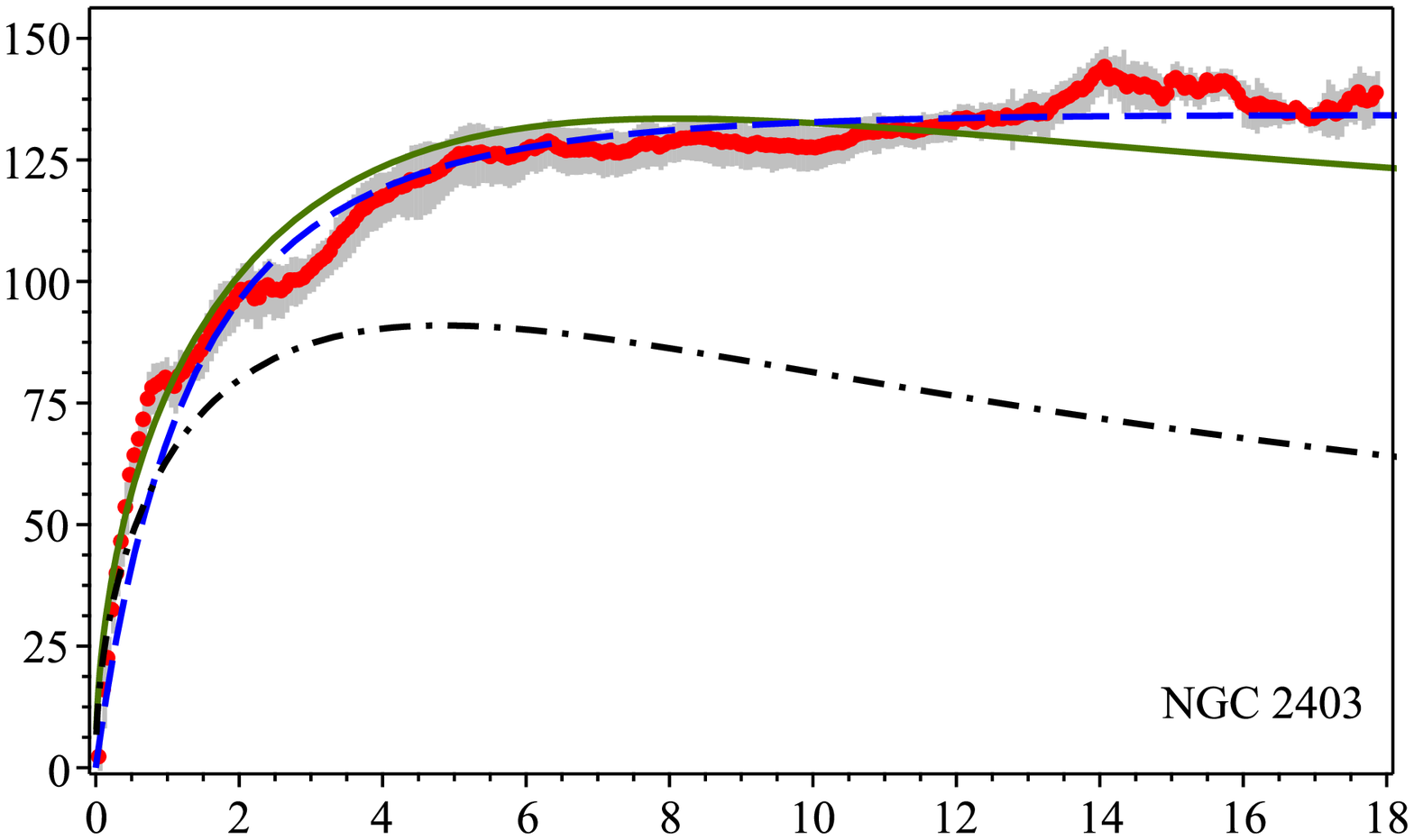}
\includegraphics[scale=0.293]{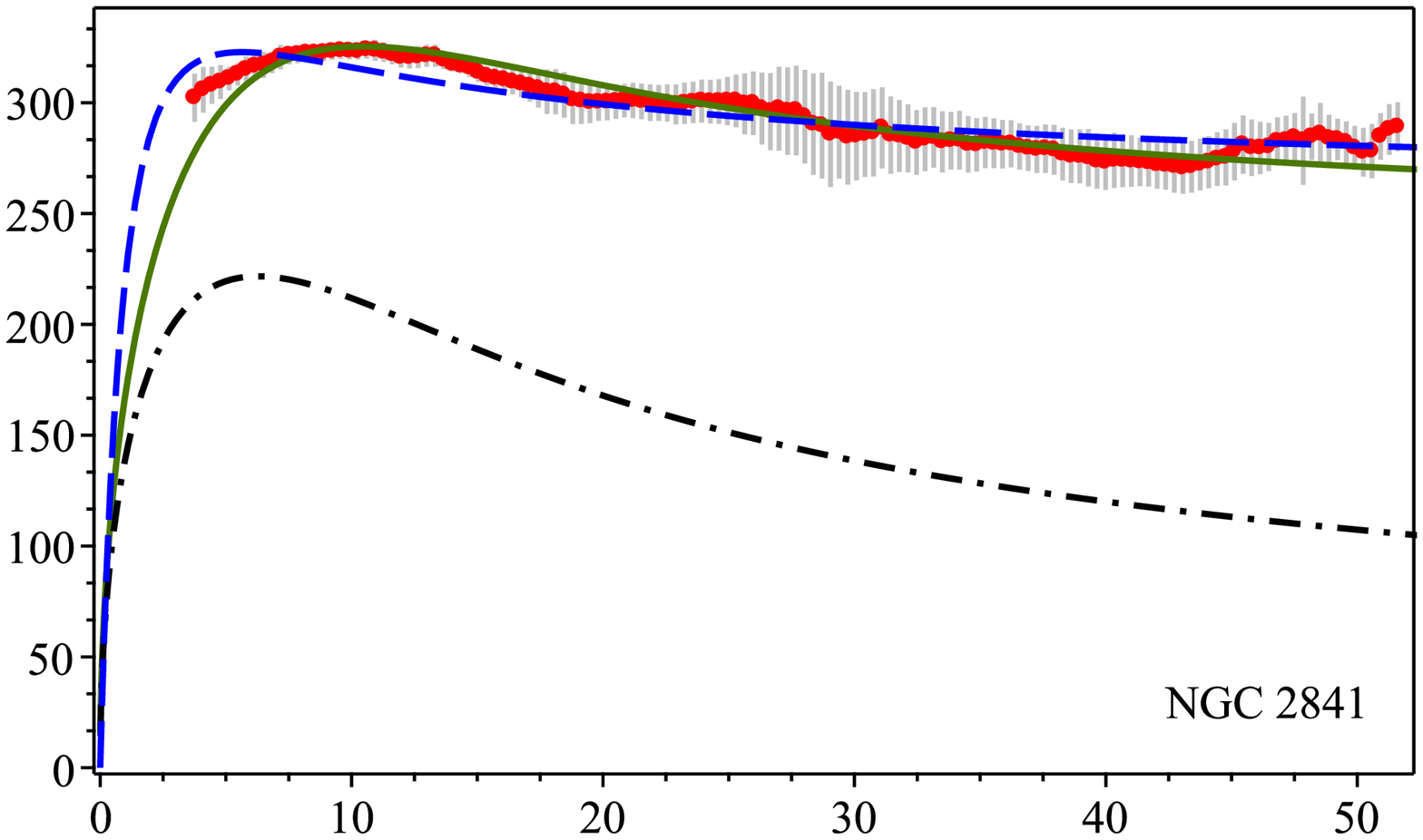}
\includegraphics[scale=0.293]{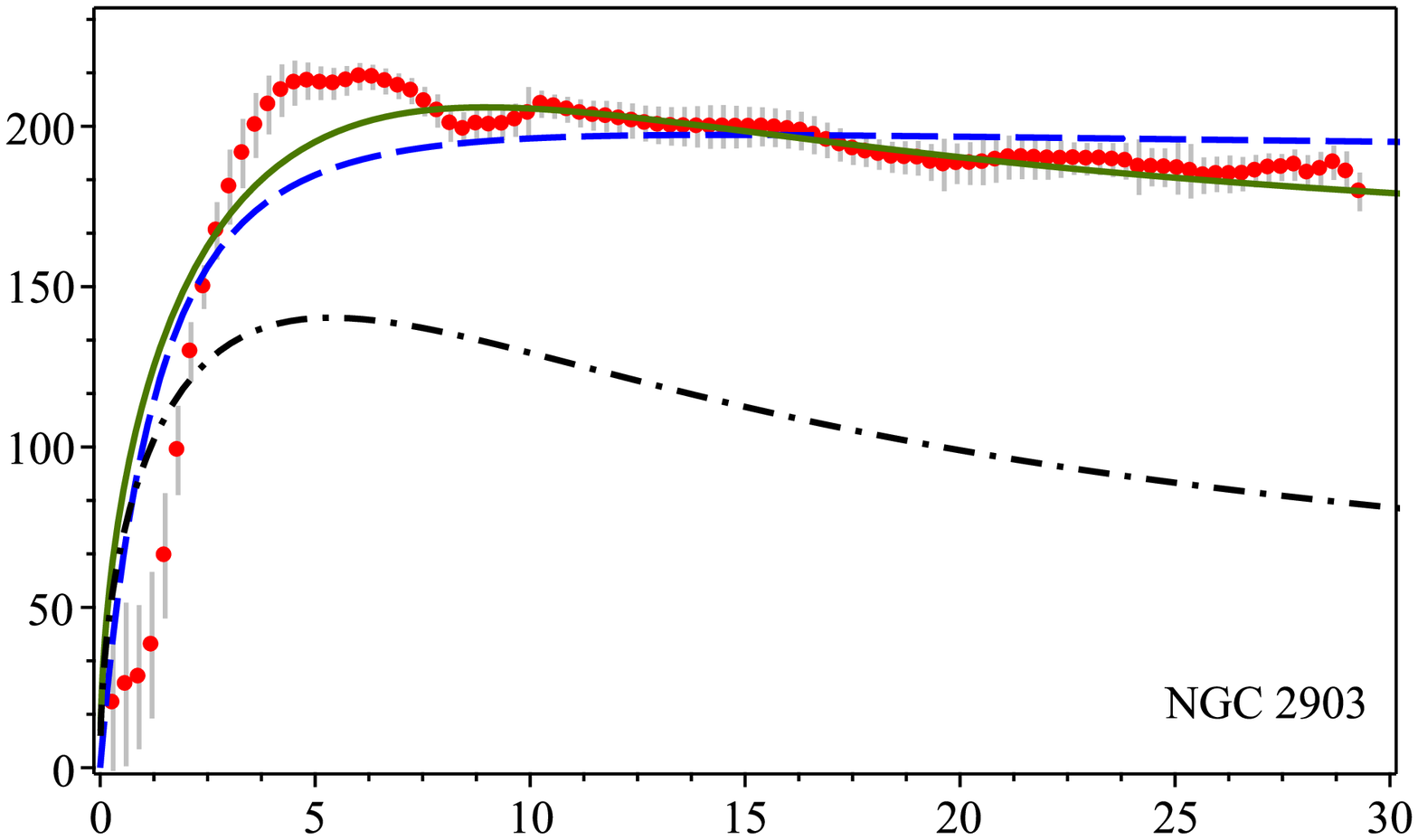}
\includegraphics[scale=0.293]{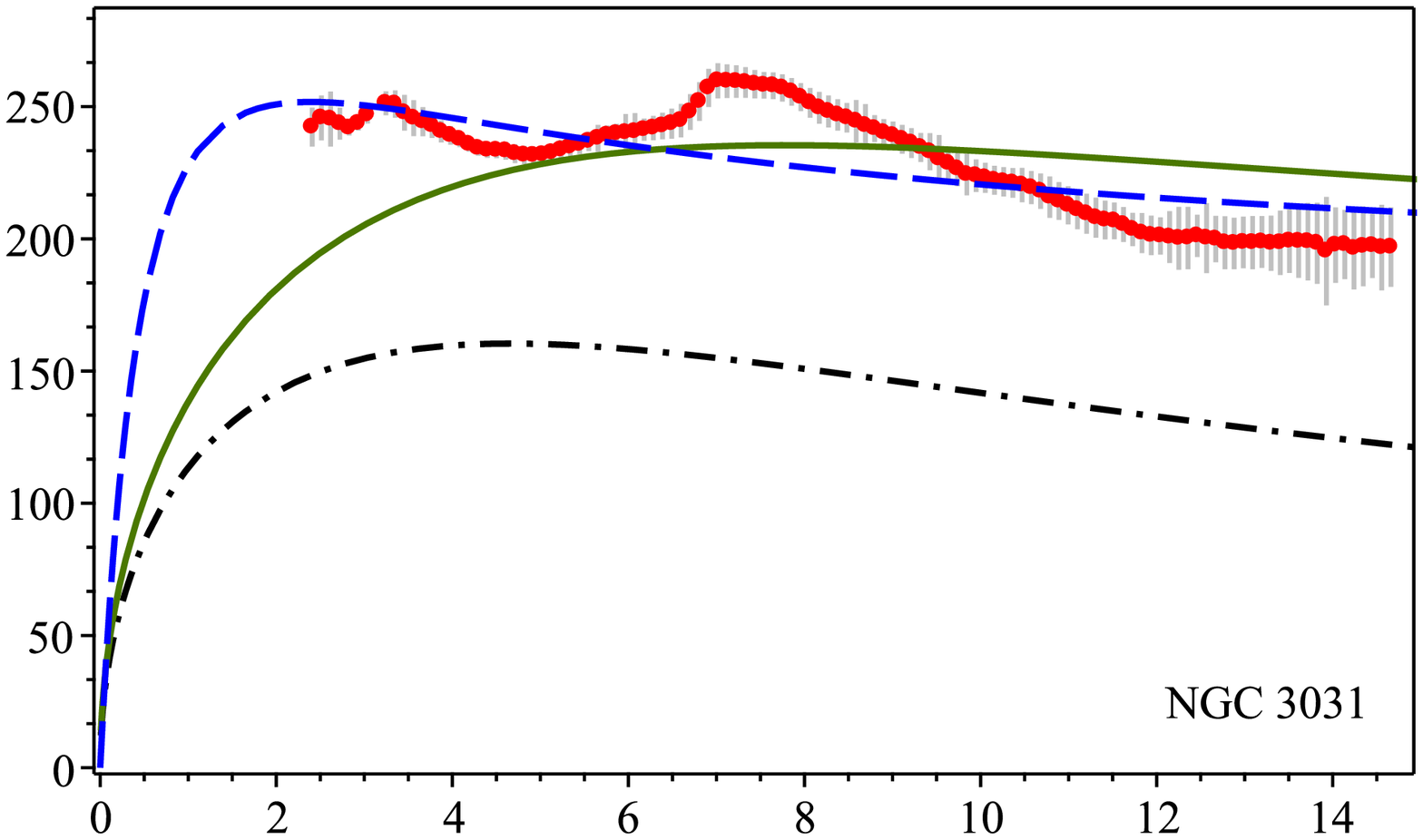}
\includegraphics[scale=0.293]{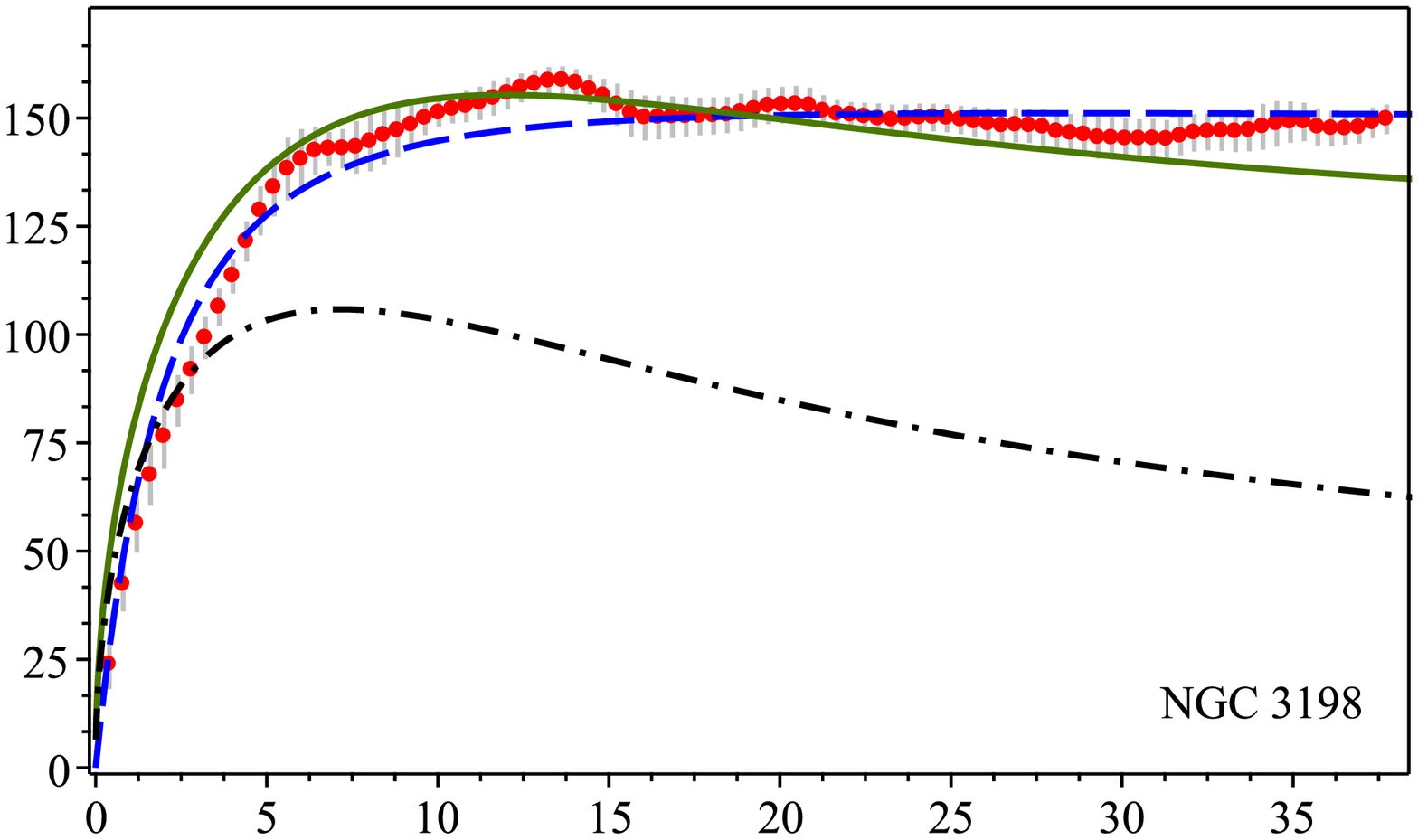}
\includegraphics[scale=0.293]{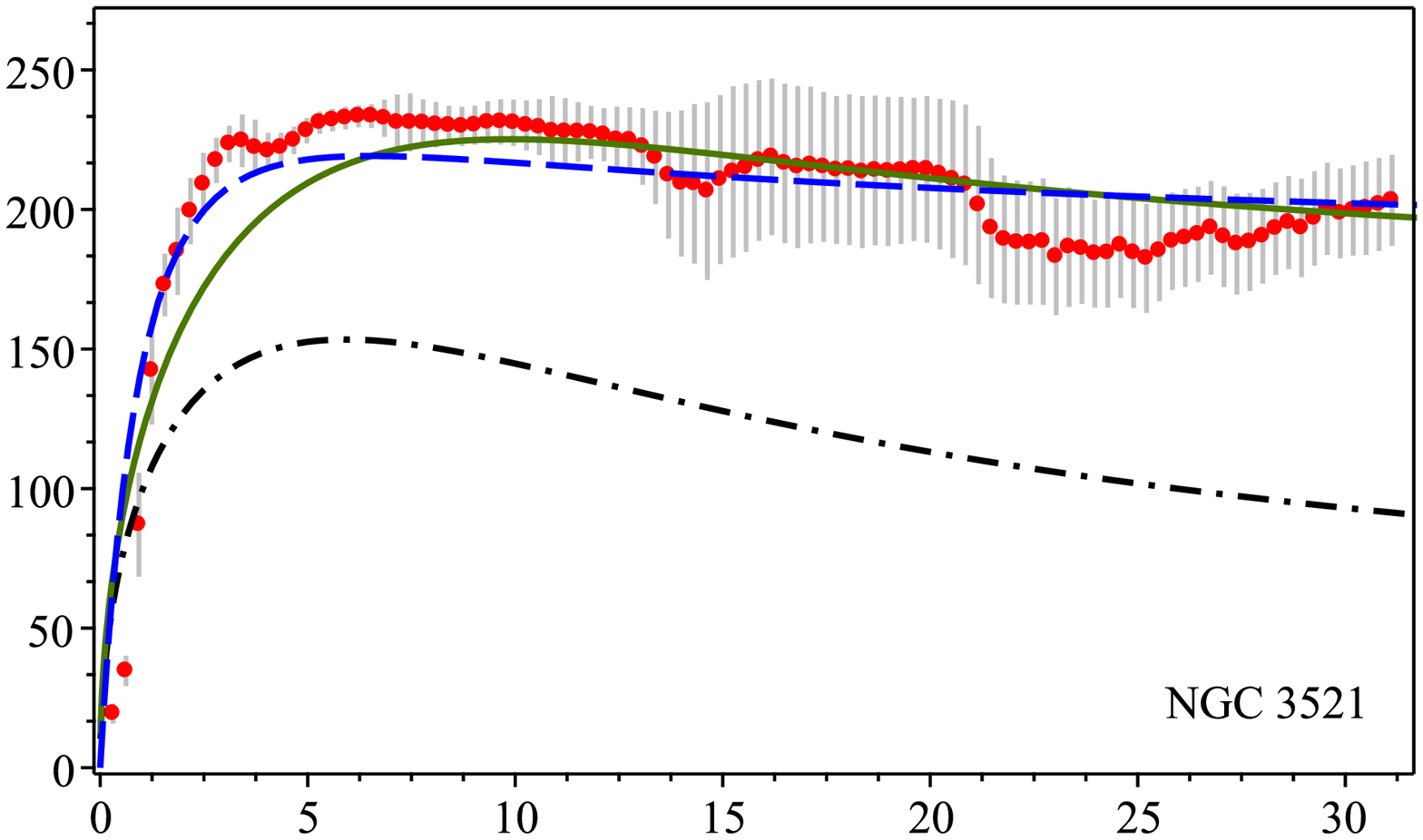}
\includegraphics[scale=0.293]{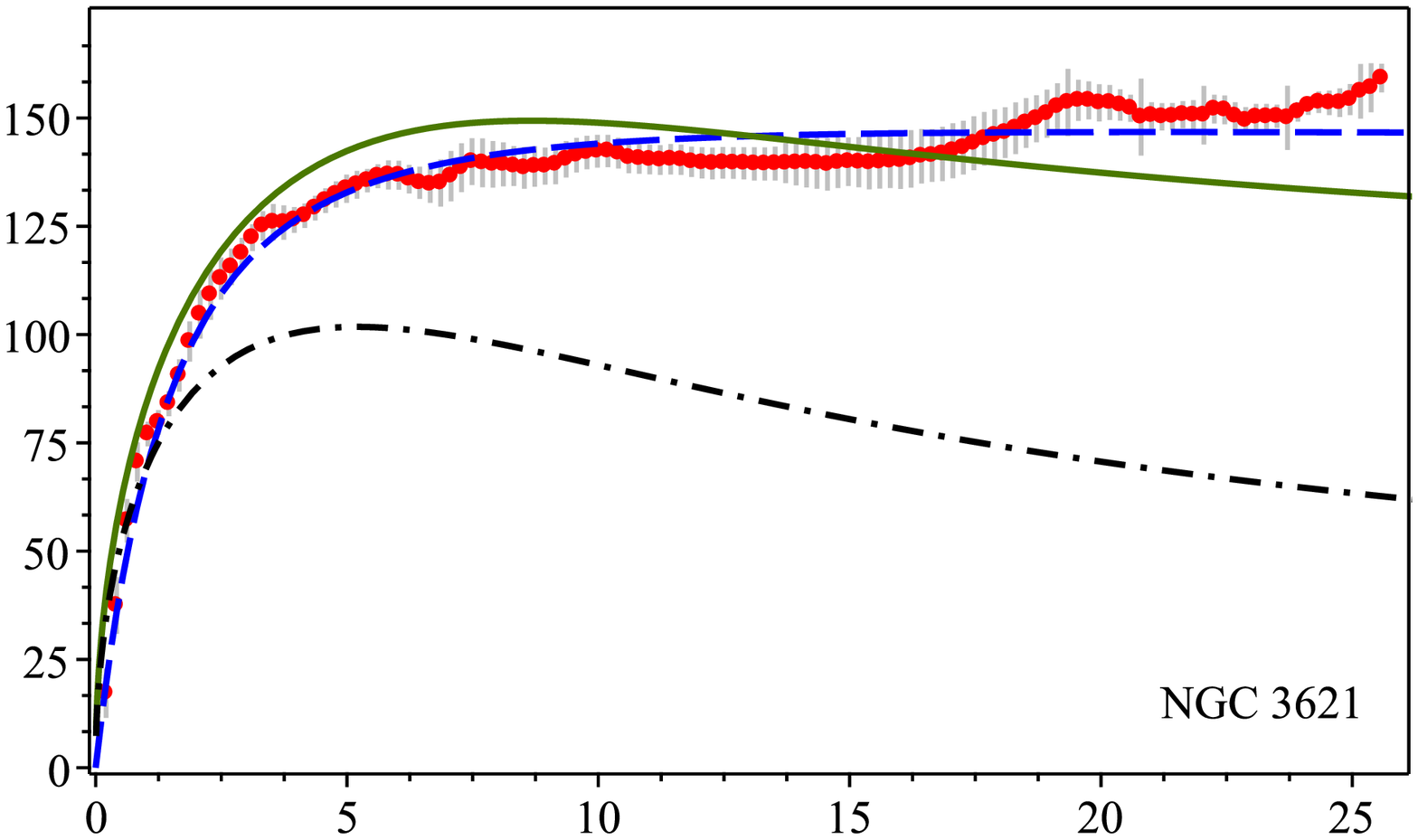}
\includegraphics[scale=0.293]{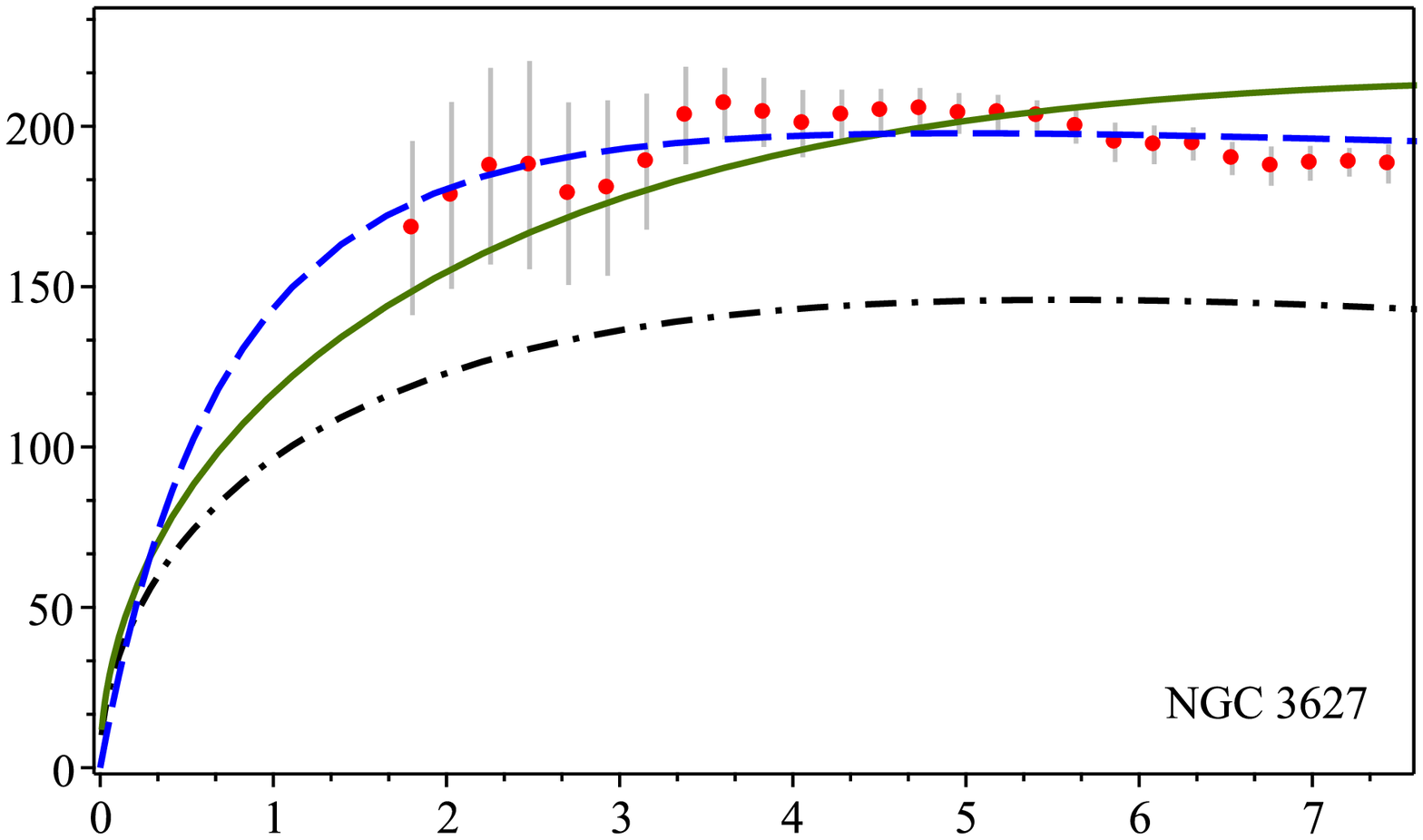}
\includegraphics[scale=0.293]{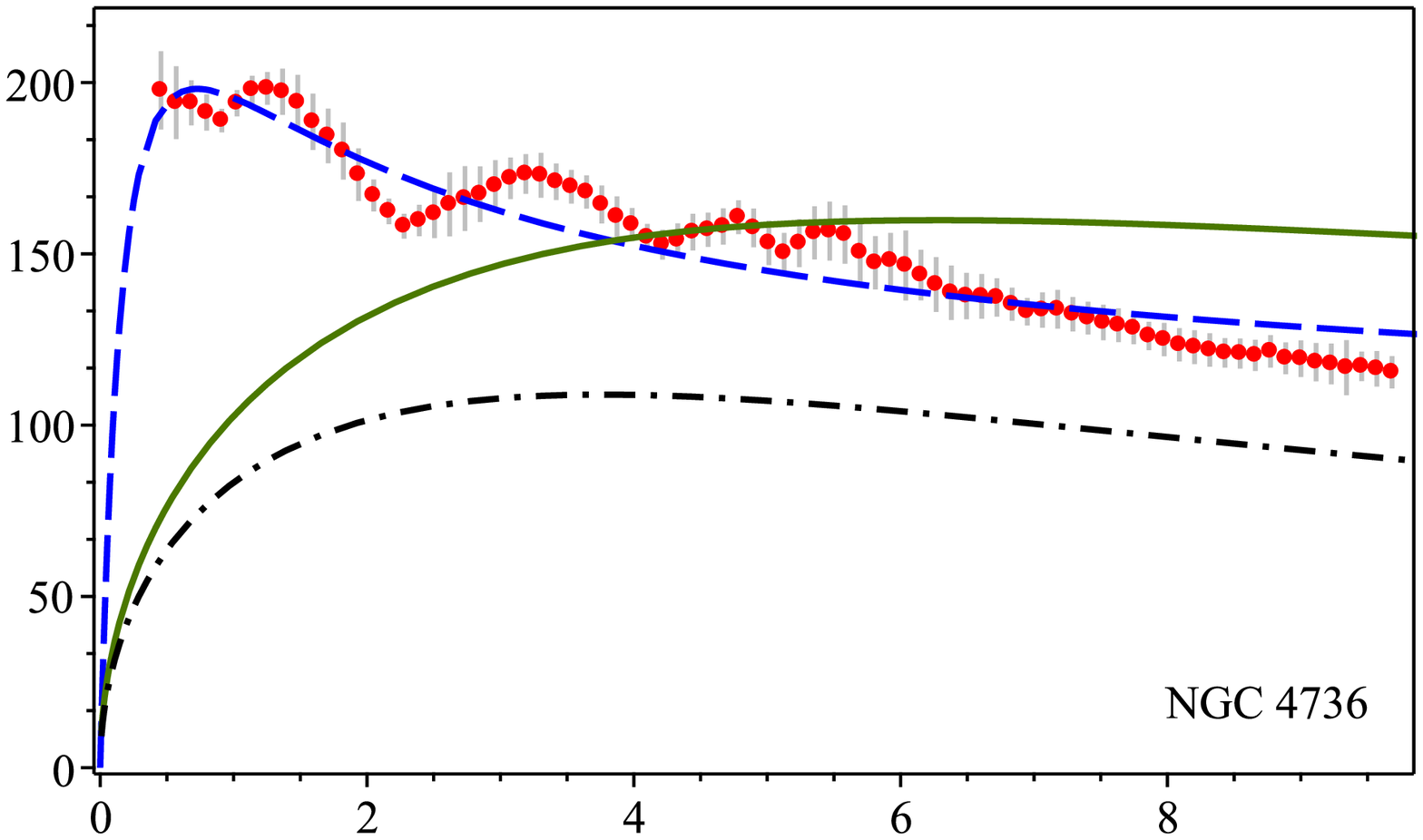}
\includegraphics[scale=0.293]{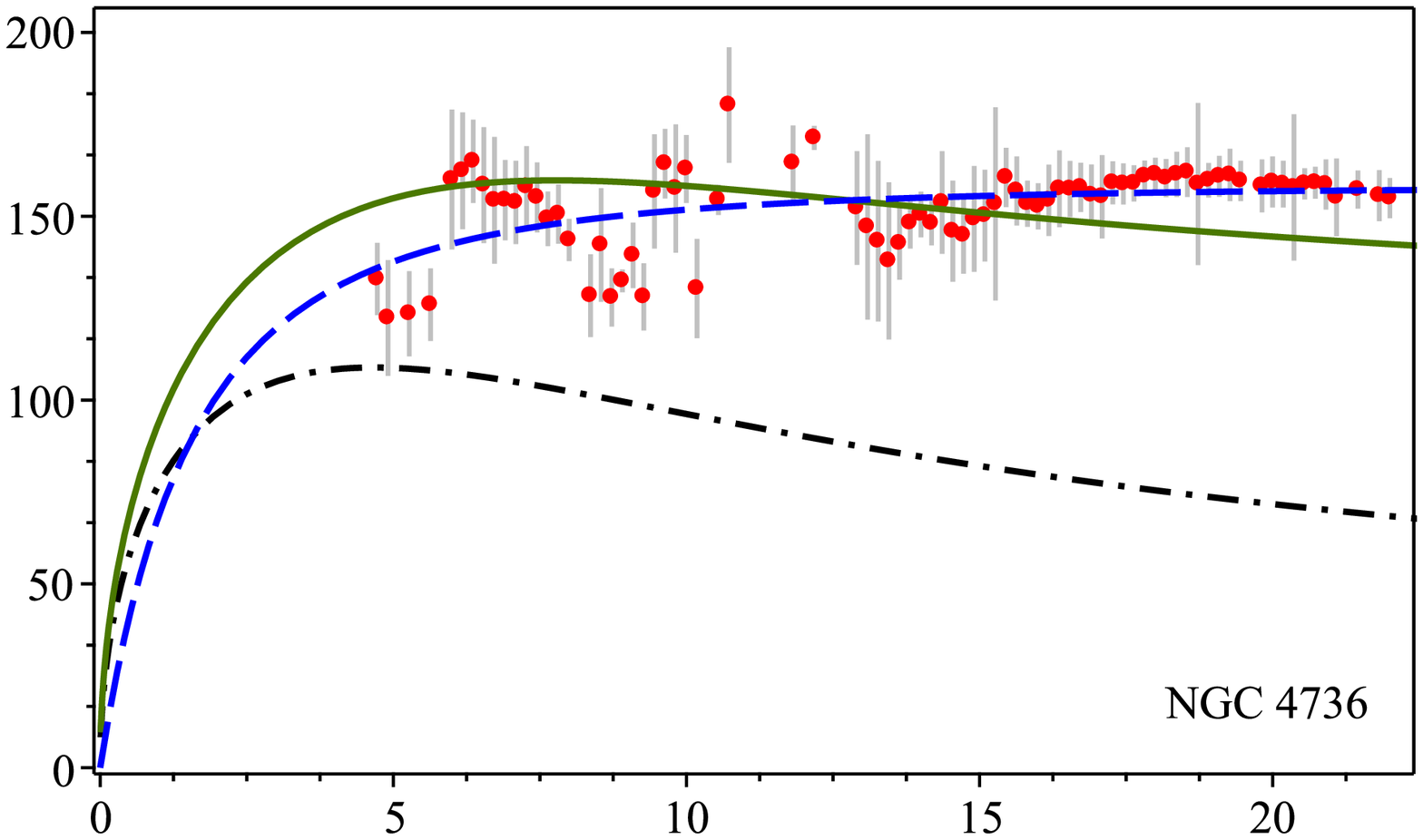}
\includegraphics[scale=0.293]{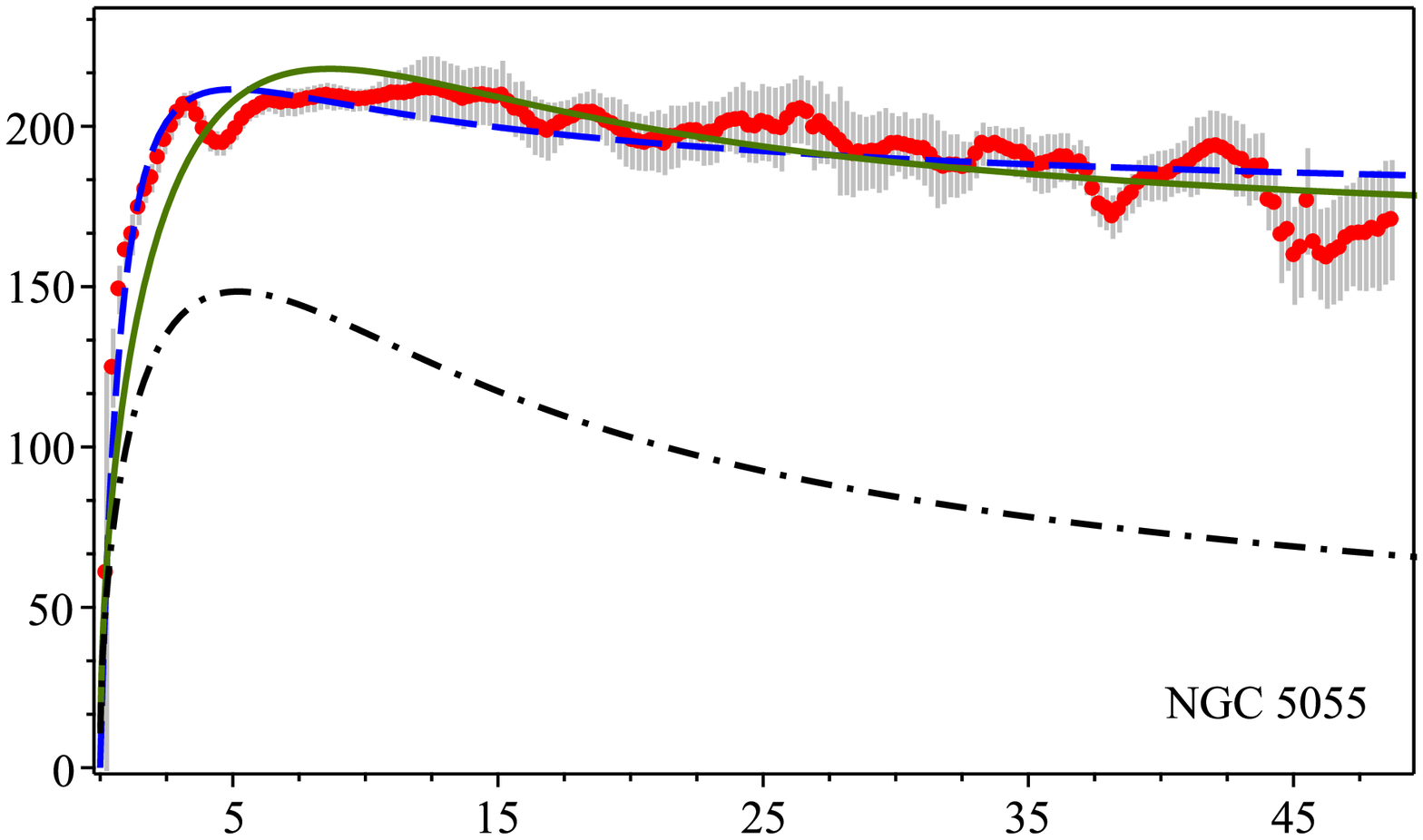}
\includegraphics[scale=0.293]{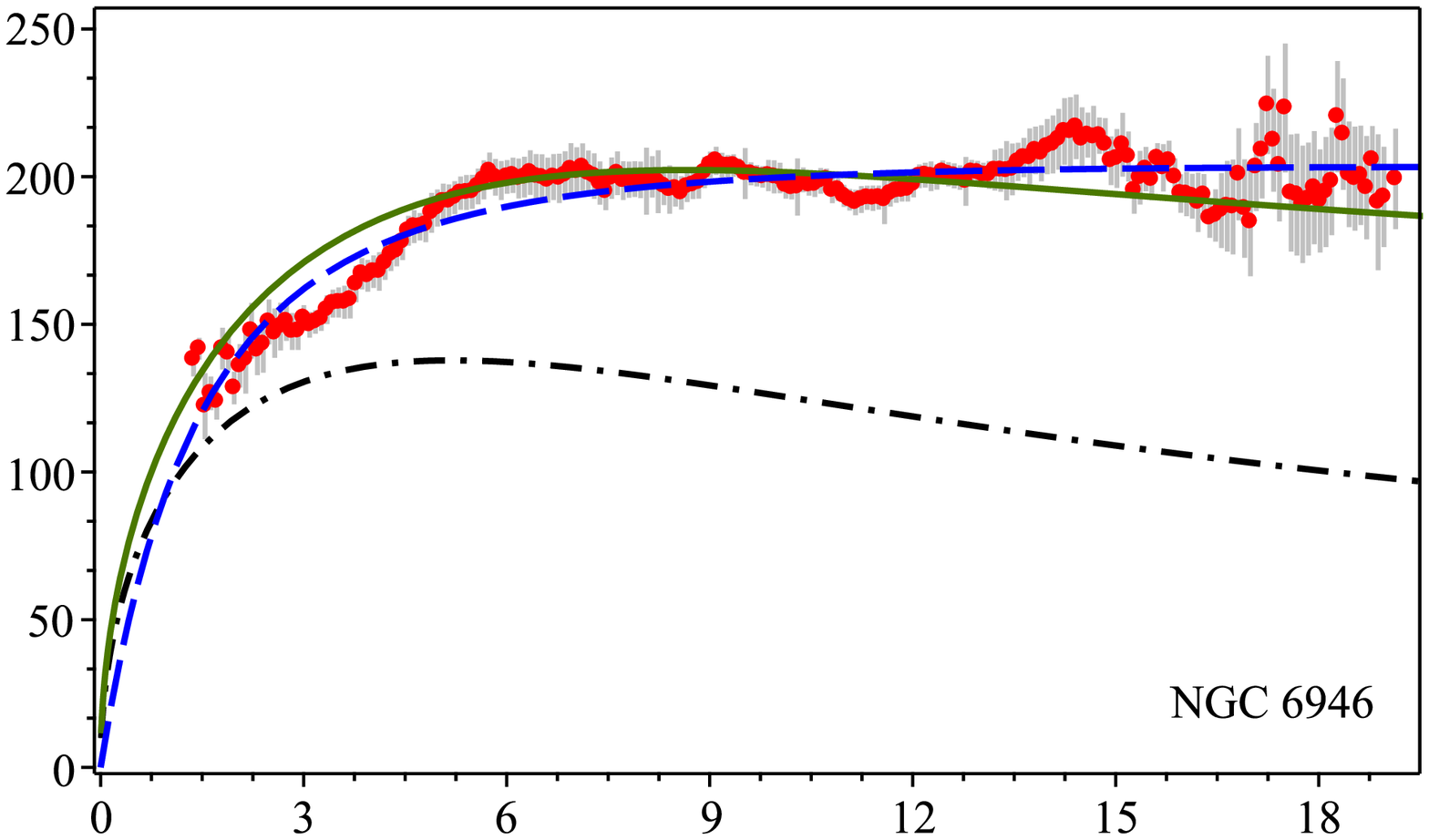}
\includegraphics[scale=0.293]{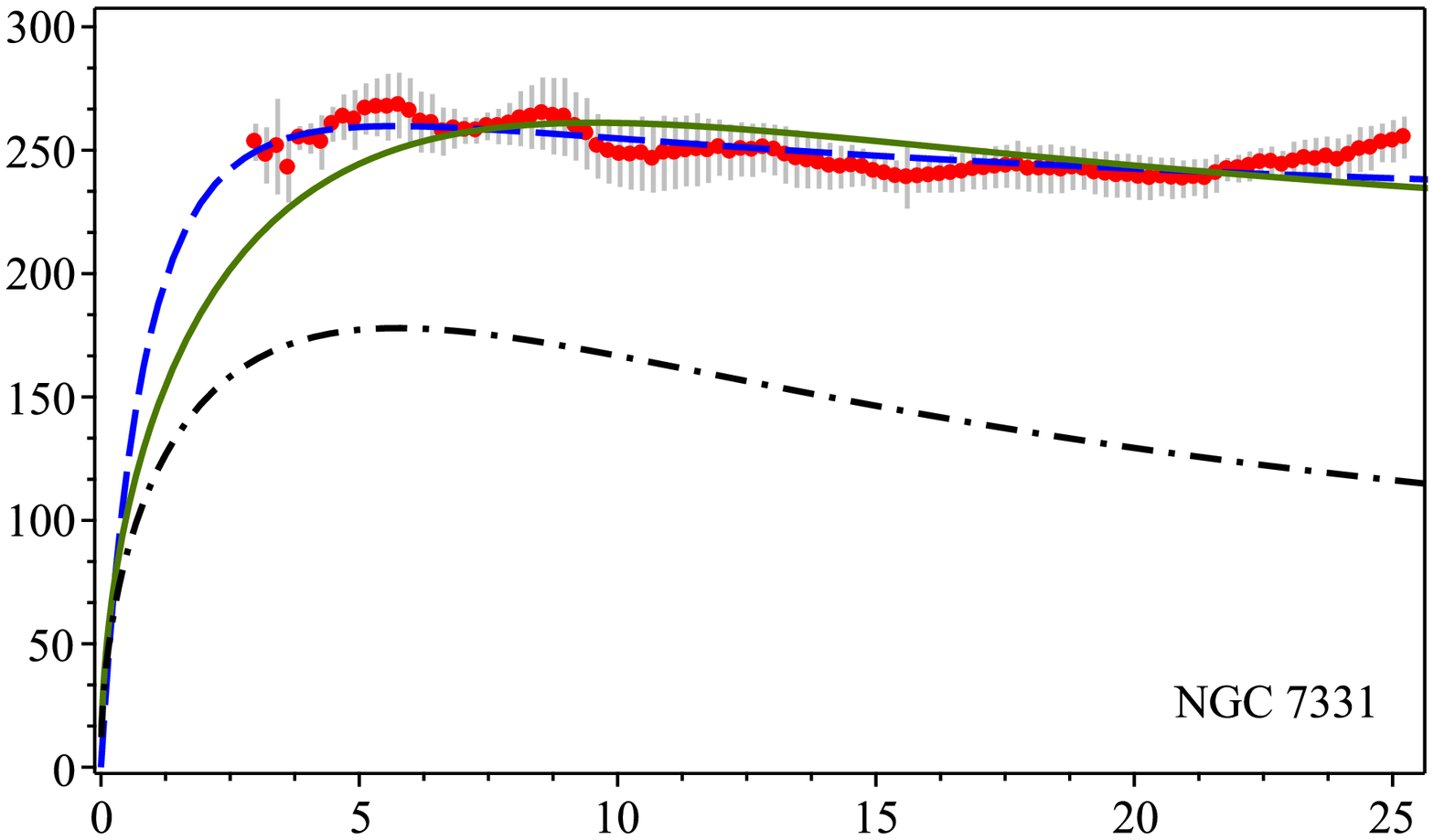}
\includegraphics[scale=0.293]{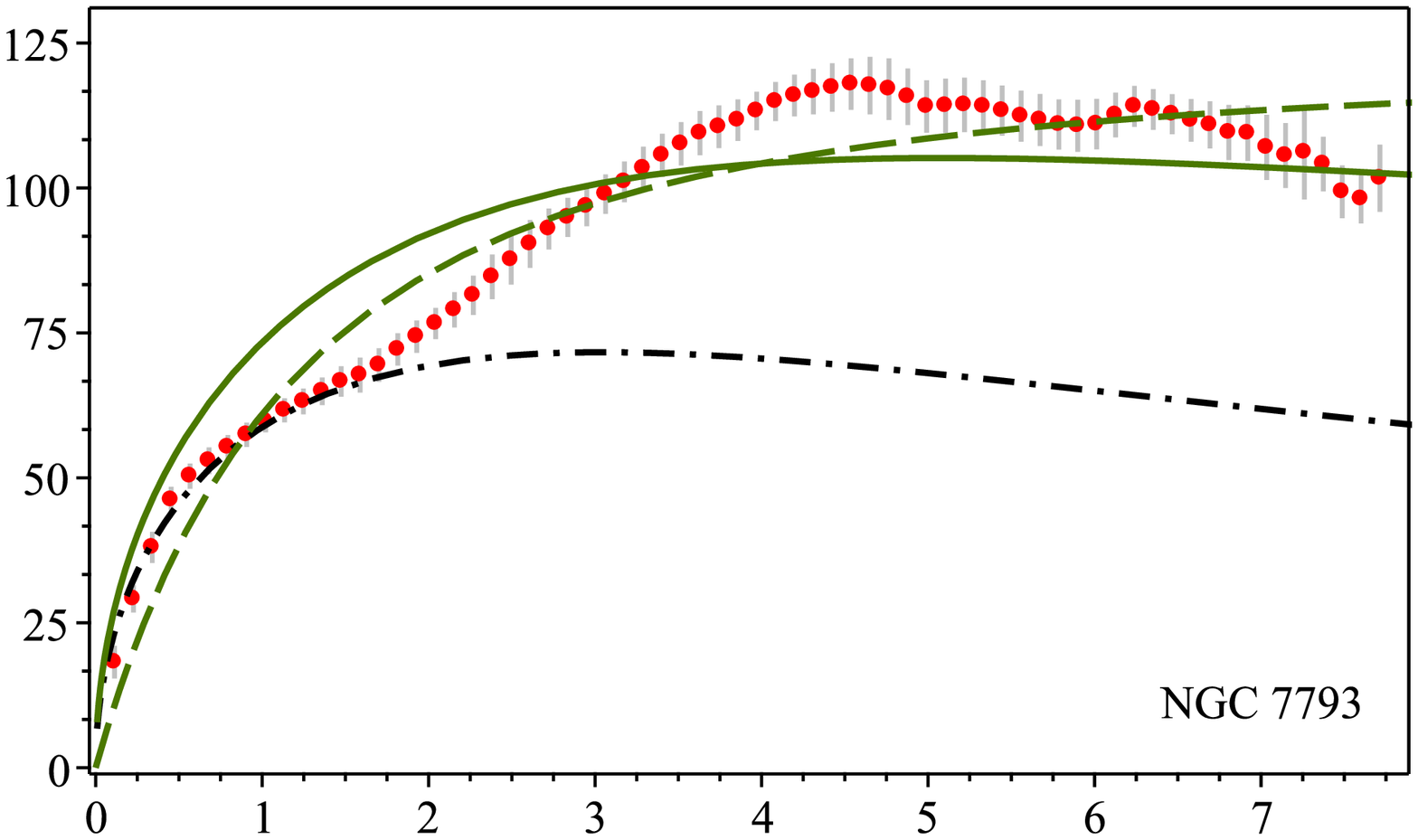}
\includegraphics[scale=0.293]{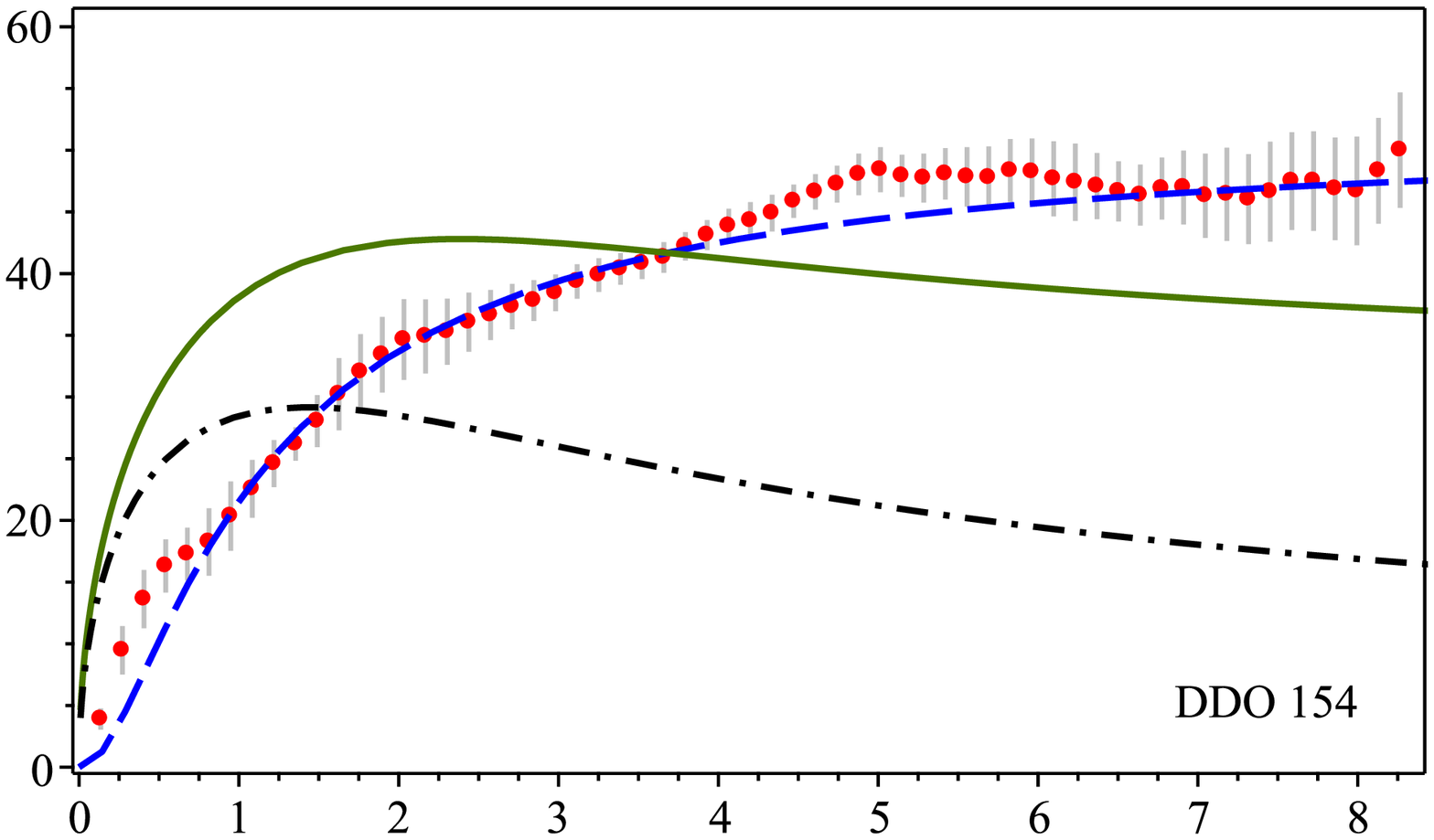}
\includegraphics[scale=0.293]{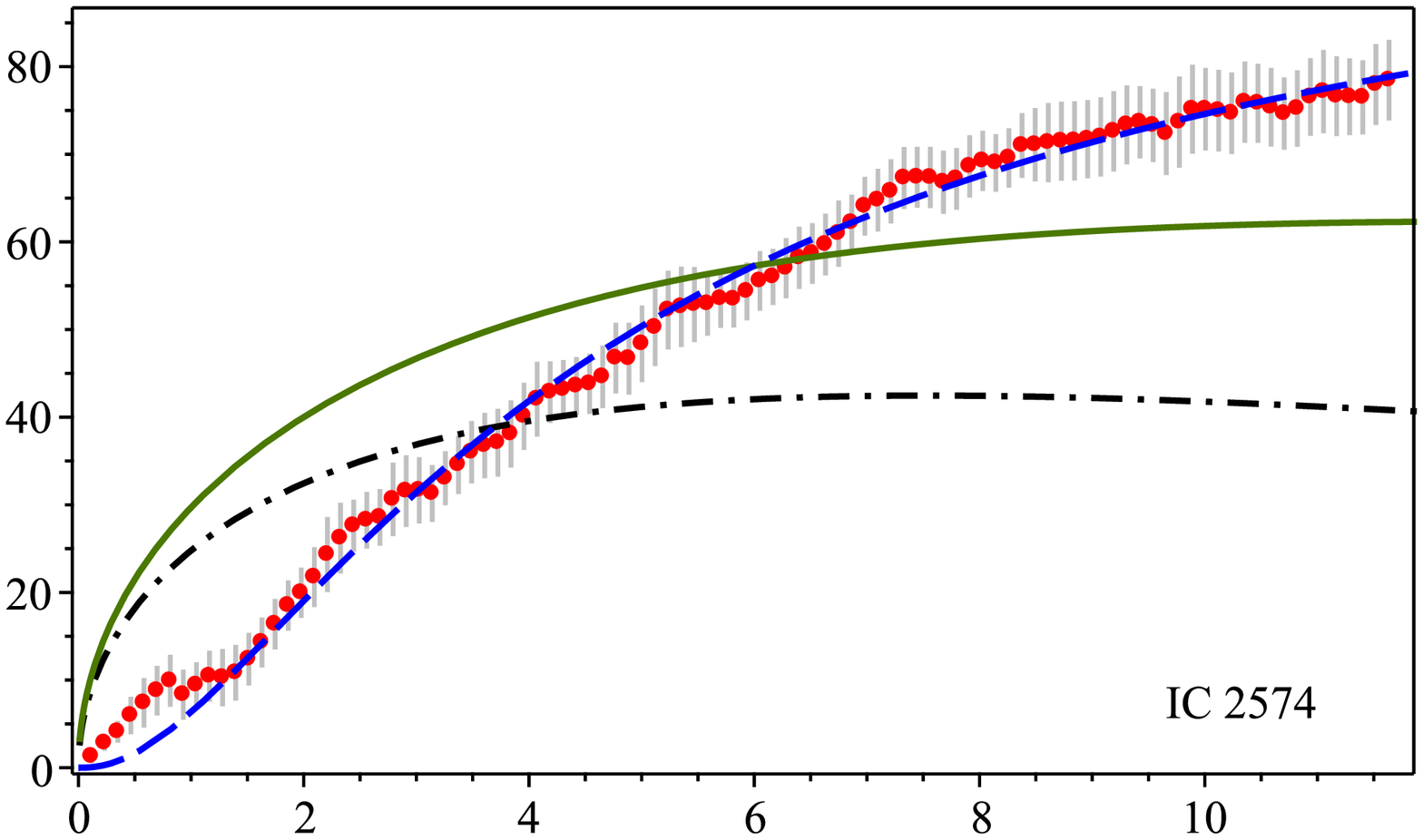}
\includegraphics[scale=0.293]{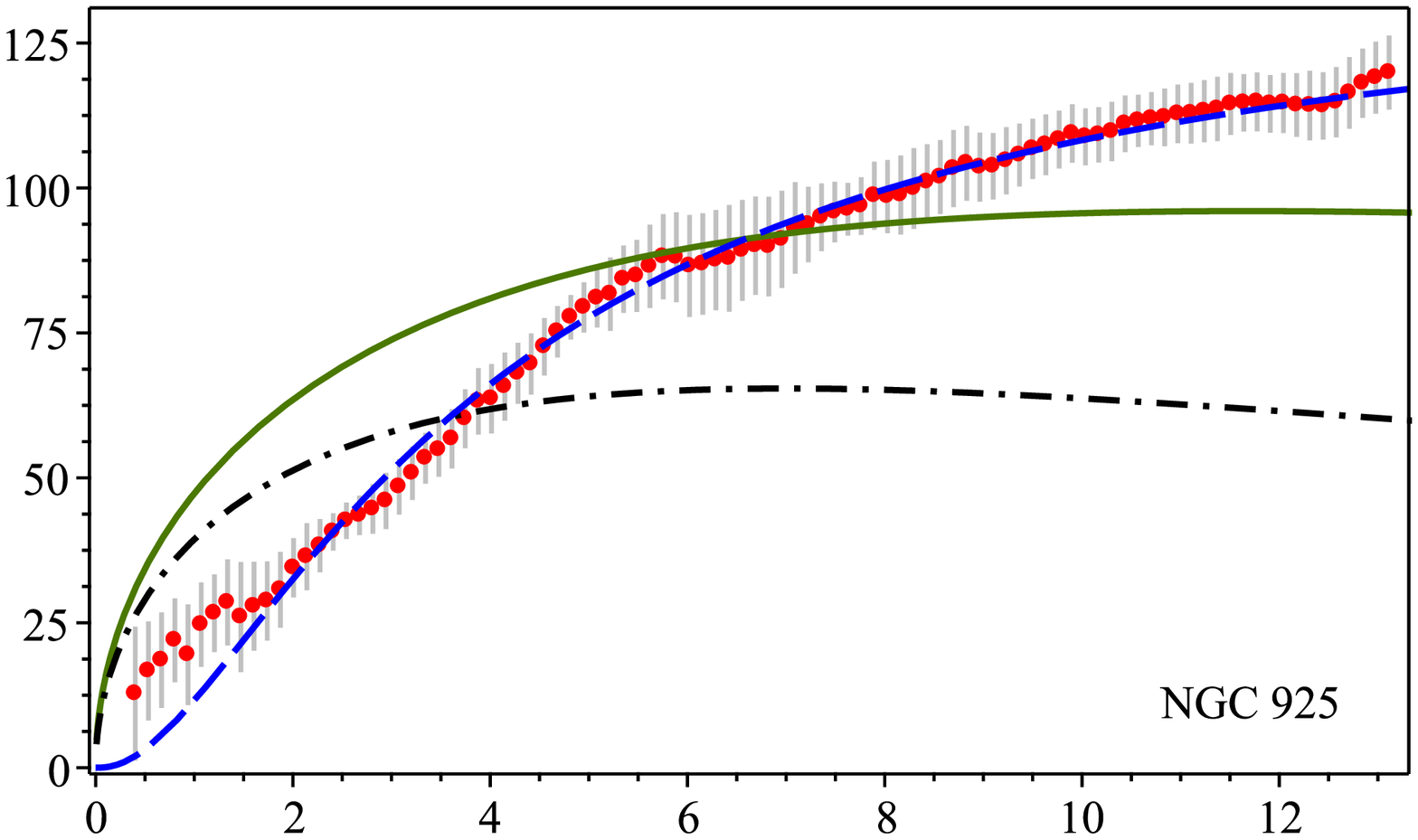}
\includegraphics[scale=0.293]{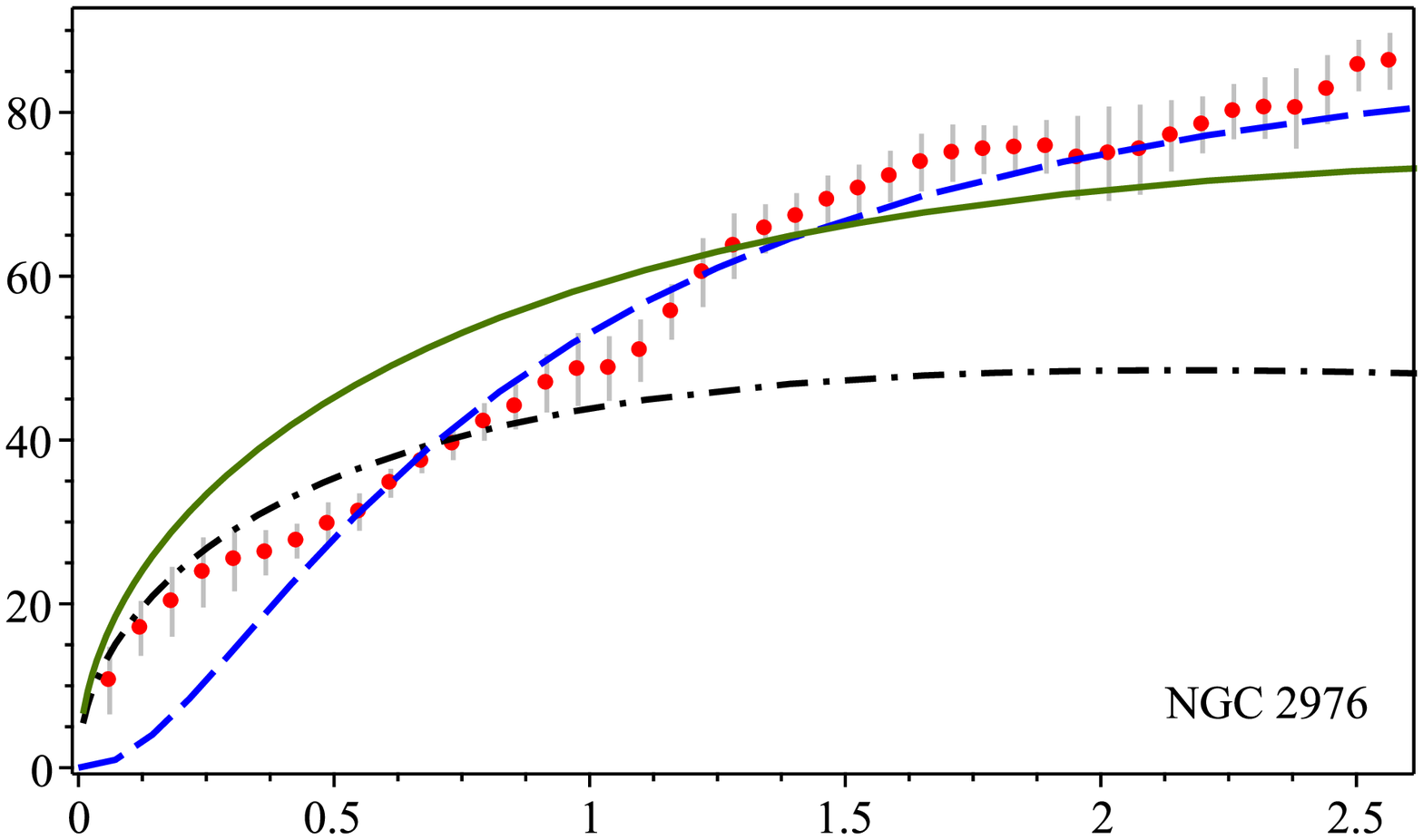}
\caption{(color online) Rotational velocities in km/s ($y$ axis) at a certain distance in kpc ($x$ axis) from the center of the galaxy. The blue curves RC are obtained from the parametric fit of eq. (\ref{vfinal}) in the case of HSB THINGS galaxies. The full (green) curve are the rotation curves obtained using
the spherical mass distribution (\ref{masssph}); the (red) full circles are the observed data points where the vertical (grey) lines represent
the error bars; the contribution due to the Newtonian term is given by the dash-dotted (black) lines, while the dashed (blue) lines give
rotation curves obtained using the mass distribution in eq. (\ref{massnew}). The numerical values resulted from the fits are given in
Table \ref{tab3}.}
\label{fig.4}
\end{figure*}

A very simple model that fits well the data (as can be seen from Figs. \ref{fig.2}, \ref{fig.4}) is obtained by choosing
\begin{equation}\label{func}
A(r)=b\left(\frac{r+r_0}{r_0}\right)
\end{equation}
where $b$ and $r_0$ are two parameters. Inserting eq. (\ref{func}) into the velocity formula (\ref{result}) we obtain
\begin{equation}\label{final}
v^2(r)=\frac{GM}{r}\left[ 1 + b\left( 1+ \frac{r}{r_0}\right) \right].
\end{equation}
In the non-relativistic limit the circular velocity and the gravitational potential are related through the usual formula $v^2(r)=r\frac{d\Phi}{dr}$, from
which it follows immediately that
\begin{equation}\label{potential}
\Phi(r)=-\frac{GM}{r}\left\{ 1+b\left[1-\frac{r}{r_0}\ln\left(\frac{r}{r_0}\right) \right] \right\}.
\end{equation}
The dependence on $\ln(r/r_0)$ in the potential was also reported in refs. \cite{mendoza, rodr, rodr2, rodr3}. Moreover, we observe that in the limit $b\rightarrow 0$ both
equations (\ref{final}) and (\ref{potential}) reduce to their usual Newtonian expressions.

Using the matter distribution (\ref{massnew}) and identifying the parameter $r_0$ contained in eq.(\ref{final}) with the galaxy scale
length $R_0$, the final rotational velocity of stars moving in circular orbits is
\begin{equation}\label{vfinal}
v^2(r)=\frac{GM_0}{r}\left( \sqrt{\frac{R_0}{r_c}}\frac{r}{r+r_c} \right)^{3\beta}\left[ 1 + b\left( 1+ \frac{r}{R_0}\right) \right]
\end{equation}
One can immediately deduce an important feature of the above formula, namely that in the limit of large radii we obtain flat
rotation curves, similar to what happens in MOND theories \cite{millg, millg2, mond1,mond2,mond3,mond4,mc1, mc2} (see also eq. (\ref{vc}) above)
\begin{equation}\label{rev2}
v_0=\sqrt{\frac{GMb}{R_0}\left(\frac{R_0}{r_c}\right)^{3\beta/2}}
\end{equation}

From the analysis of the 18 THINGS galaxies sample we have found $b=0.352\pm0.08$ to give a good fit results for the rotation
curves. The plots in Fig. \ref{fig.2} and the best fit results from Table \ref{tab2} are obtained using the value $b=0.352$. As
in \cite{Moffat2} the value $\beta=1$ (for HSB galaxies) and $\beta=2$ (LSB galaxies) give good fit results. By allowing $\beta$ to be a free parameter, slightly better fits results can be obtain. In
this case a preliminary analysis indicates that $0.75<\beta<1.25$ for HSB galaxies and $1.9<\beta<2.1$ for LSB galaxies. However, in order
to keep the free parameters to a minimum we have chosen here to fix the value of $\beta$.

\begin{table*}
\renewcommand{\arraystretch}{1.3}
\caption{\label{tab:table3} Best fit results according to eq.(\ref{vfinal}) using the parametric mass distribution (\ref{massnew}). These numerical values correspond
to rotation curves presented in Fig. \ref{fig.2}. Col. (1) name of galaxy; col (2) distance; col. (3) measured scale length of the galaxy; col. (4) base ten logarithm of
total gas mass given by $M_{gas}=4/3M_{HI}$, with the $M_{HI}$ data taken from \cite{Walter}; col. (5) galaxy luminosity in the B-band calculated from \cite{Walter};
col. (6) base ten logarithm of the predicted stelar mass $M_*$ of the galaxy (obtained by subtracting $M_{gas}$ from the best-fit results for the total mass $M_0$);
col. (7) the predicted core radius $r_c$; col. (8) reduced $\chi^2_r$; col. (9) the stelar mass-to-light ratio calculated by subtracting the mass of the gas from the
total mass and then dividing it by the B-band luminosity; col. (10) base ten logarithm of MOND predicted mass of the galaxy; col. (11) the MOND predicted core
radius $r_c$; col. (12) MOND reduced $\chi^2_r$; and col. (13) the MOND stelar mass-to-light ratio.}
\label{tab2}
\begin{tabular*}{\textwidth}{crcrcrrrcrcrrc}
\hline
  & & & & & & & & & & \multicolumn{1}{r}{MOND}  \\ \cline{10-13}
  Galaxy & \multicolumn{1}{c}{$D$} & $R_0$ & \multicolumn{1}{c}{$\log M_{gas}$} & $L_B$ & \multicolumn{1}{c}{$\log M_*$} & \multicolumn{1}{c}{$r_c$} & $\chi^2_r$ & \multicolumn{1}{c}{$M_*/L$} & \multicolumn{1}{c}{$\log M_*$} & \multicolumn{1}{c}{$r_c$} & $\chi^2_r$ & $M_*/L$ \\
  & \multicolumn{1}{c}{Mpc} & kpc & \multicolumn{1}{c}{$M_{\odot}$} &\multicolumn{1}{c}{$10^{10}\,L_{\odot}$} & \multicolumn{1}{c}{$M_{\odot}$}& \multicolumn{1}{c}{kpc} &  & \multicolumn{1}{c}{$M_{\odot}/L_{\odot}$} & \multicolumn{1}{c}{$M_{\odot}$} & \multicolumn{1}{c}{kpc} & & \multicolumn{1}{c}{$M_{\odot}/L_{\odot}$} & \\
  (1) & \multicolumn{1}{c}{(2)} & \multicolumn{1}{c}{(3)} & \multicolumn{1}{c}{(4)} & \multicolumn{1}{c}{(5)} & \multicolumn{1}{c}{(6)} & \multicolumn{1}{c}{(7)} & \multicolumn{1}{c}{(8)} & \multicolumn{1}{c}{(9)} & \multicolumn{1}{c}{(10)} & (11) & (12) & (13) \\ \hline
HSB type \\ \hline
 NGC 2403 &  3.2 & 2.7 &  9.53 & 0.921 & 10.36 & 2.48 & 1.32 & 2.49 & 10.21 & 2.06 & 0.69 & 1.78 \\
 NGC 2841 & 14.1 & 3.5 & 10.06 & 4.742 & 10.64 & 1.73 & 1.11 & 0.92 & 11.50 & 2.81 & 1.71 & 6.71 \\
 NGC 2903 &  8.9 & 3.0 &  9.76 & 3.664 & 10.67 & 2.51 & 5.30 & 1.29 & 11.06 & 2.85 & 7.94 & 3.14 \\
 NGC 3031 &  3.6 & 2.6 &  9.68 & 3.049 &  9.76 & 0.88 & 5.63 & 0.19 & 10.66 & 1.37 & 6.07 & 1.52 \\
 NGC 3198 & 13.8 & 4.0 & 10.13 & 3.106 & 10.58 & 3.76 & 1.61 & 1.23 & 10.19 & 2.96 & 3.99 & 0.50 \\
 NGC 3521 & 10.7 & 3.3 & 10.03 & 3.698 & 10.31 & 1.84 & 5.19 & 0.55 & 10.78 & 2.09 & 6.31 & 1.65 \\
 NGC 3621 &  6.6 & 2.9 &  9.97 & 1.629 & 10.42 & 2.75 & 1.49 & 1.63 & 10.28 & 2.29 & 0.85 & 1.17 \\
 NGC 3627 &  9.3 & 3.1 &  9.04 & 3.076 & 10.23 & 1.53 & 0.83 & 0.56 & 10.68 & 1.87 & 0.91 & 1.59 \\
 NGC 4736 &  4.7 & 2.1 &  8.72 & 1.294 &  8.42 & 0.32 & 2.50 & 0.02 &  8.93 & 0.34 & 5.18 & 0.07 \\
 NGC 4826 &  7.5 & 2.6 &  8.86 & 2.779 & 10.67 & 2.85 & 1.57 & 1.71 & 10.61 & 2.27 & 1.61 & 1.46 \\
 NGC 5055 & 10.1 & 2.9 & 10.08 & 4.365 &  9.98 & 1.50 & 1.24 & 0.22 & 10.35 & 1.47 & 2.54 & 0.51 \\
 NGC 6946 &  5.9 & 2.9 &  9.74 & 2.729 & 10.80 & 2.74 & 1.52 & 2.31 & 11.26 & 3.29 & 1.61 & 6.70 \\
 NGC 7331 & 14.7 & 3.2 & 10.08 & 7.244 & 10.40 & 1.68 & 0.37 & 0.35 & 11.13 & 2.31 & 0.24 & 1.86 \\
 NGC 7793 &  3.9 & 1.7 &  9.07 & 0.511 & 10.32 & 2.09 & 4.65 & 4.13 & 10.53 & 2.29 & 4.23 & 6.73 \\ \hline
LSB type \\ \hline
 DDO 154  & 4.3 & 0.8 & 8.68 & 0.007 & 8.62 & 0.69 & 1.01 & 6.00 &  8.34 & 0.83 & 0.59 &  3.14 \\
 IC 2574  & 4.0 & 4.2 & 9.29 & 0.273 & 9.98 & 3.07 & 0.52 & 3.49 & 10.49 & 4.86 & 0.30 & 11.40 \\
 NGC 925  & 9.2 & 3.9 & 9.78 & 1.614 & 9.34 & 2.62 & 0.31 & 0.54 & 10.55 & 3.61 & 0.25 &  2.22 \\
 NGC 2976 & 3.6 & 1.2 & 8.27 & 0.201 & 8.71 & 0.58 & 2.19 & 0.26 &  9.41 & 0.75 & 1.37 &  1.30 \\
 \hline
\end{tabular*}
\end{table*}

\begin{table}
\renewcommand{\arraystretch}{1.3}
\caption{\label{tab:table3} Best fitting results using eqs. (\ref{final}) and (\ref{masssph}). The corresponding rotation curves are given in Fig.\ref{fig.4}. Col. (3) gives best-fit results for the predicted galaxy stelar mass; col. (4) gives the values of
reduced $\chi^2_r$; and col. (5) gives the stelar mass-to-light ratio. }
\label{tab3}
\begin{tabular*}{\textwidth}{ccccccc}
\cline{2-6}
  &  Galaxy & Type &\multicolumn{1}{c}{$\log M$} & \multicolumn{1}{c}{$\chi^2_r$}  & \multicolumn{1}{c}{$M/L$} \\
  & & & \multicolumn{1}{c}{($\,M_{\odot}$)} &   & \multicolumn{1}{c}{($M_{\odot}/L_{\odot}$)} \\
  & (1) & \multicolumn{1}{c}{(2)} & \multicolumn{1}{c}{(3)} & \multicolumn{1}{c}{(4)} & \multicolumn{1}{c}{(5)} \\ \cline{2-6}
 & NGC 2403 & HSB & 10.15 & 3.88 & 1.51 \\
 & NGC 2841 & HSB & 11.08 & 0.55 & 2.58 \\
 & NGC 2903 & HSB & 10.60 & 2.10 & 1.09 \\
 & NGC 3031 & HSB & 10.67 & 15.01 & 1.55 \\
 & NGC 3198 & HSB & 10.33 & 3.55 & 0.69 \\
 & NGC 3521 & HSB & 10.69 & 6.18 & 1.34 \\
 & NGC 3621 & HSB & 10.14 & 10.41 & 0.86 \\
 & NGC 3627 & HSB & 10.70 & 4.94 & 1.63 \\
 & NGC 4736 & HSB & 10.27 & - & 1.46 \\
 & NGC 4826 & HSB & 10.36 & 3.14 & 0.83 \\
 & NGC 5055 & HSB & 10.57 & 2.99 & 0.86 \\
 & NGC 6946 & HSB & 10.57 & 2.55 & 1.37 \\
 & NGC 7331 & HSB & 10.82 & 1.84 & 0.92 \\
 & NGC 7793 & HSB & 9.75  & 12.47 & 1.09 \\
 & DDO 154 & LSB & 7.69 & 21.17 & 0.71 \\
 & IC 2574 & LSB & 9.59 & 18.47 & 1.43 \\
 & NGC 925 & LSB & 9.83 & 10.98 & 0.42 \\
 & NGC 2976 & LSB & 9.34 & 12.98 & 1.10 \\
 \cline{2-6}
\end{tabular*}
\end{table}

If we replace the matter distribution (\ref{massnew}) in the equation (\ref{final}) with the one coming from the spherical version of the exponential
disc profile \cite{Binney}
\begin{equation}\label{masssph}
M(r)=M_0\left[ 1- \left(1+\frac{r}{R_0} \right)\exp{\left(-\frac{r}{R_0}\right)} \right],
\end{equation}
we can then fit the rotation curves using only $M/L$ as a free parameter. The resulted predicted values for the stellar mass of the galaxies are
given in Table \ref{tab3} together with the corresponding rotation curves in Fig. \ref{fig.4}.

\subsubsection{The Tully-Fisher relation}

\begin{figure*}[h!t]
\centering
\includegraphics[scale=0.4]{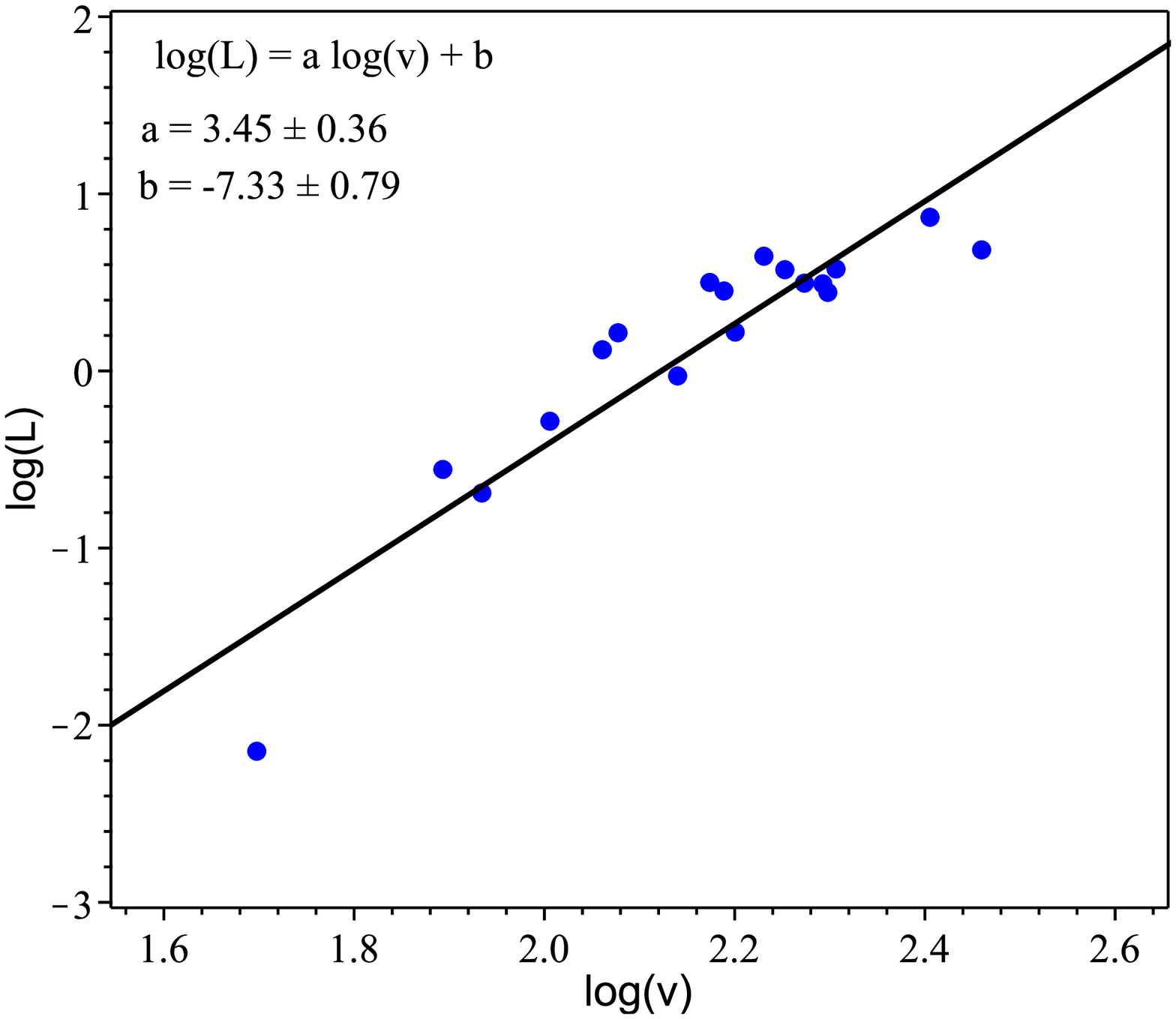}
\includegraphics[scale=0.4]{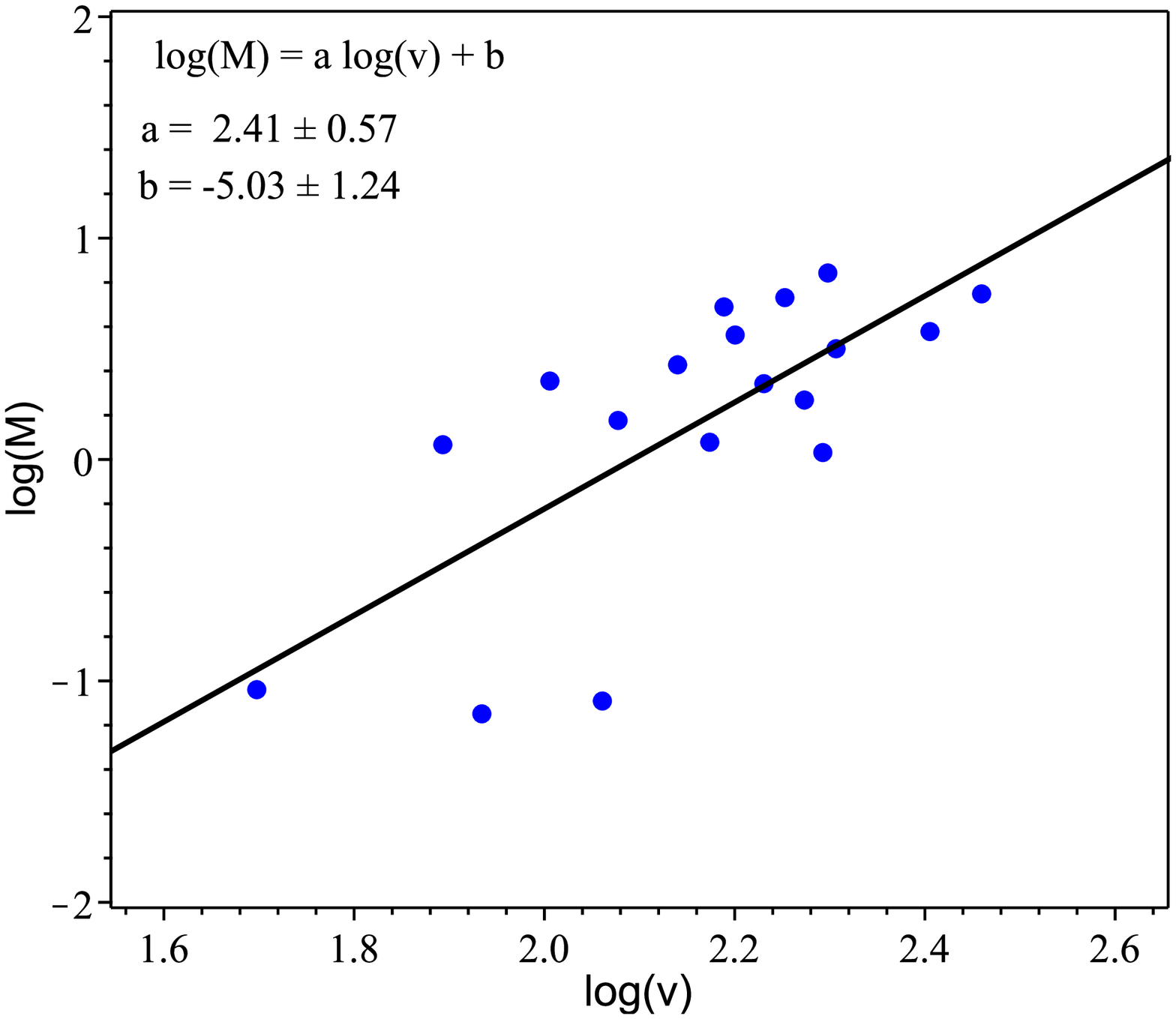}
\includegraphics[scale=0.4]{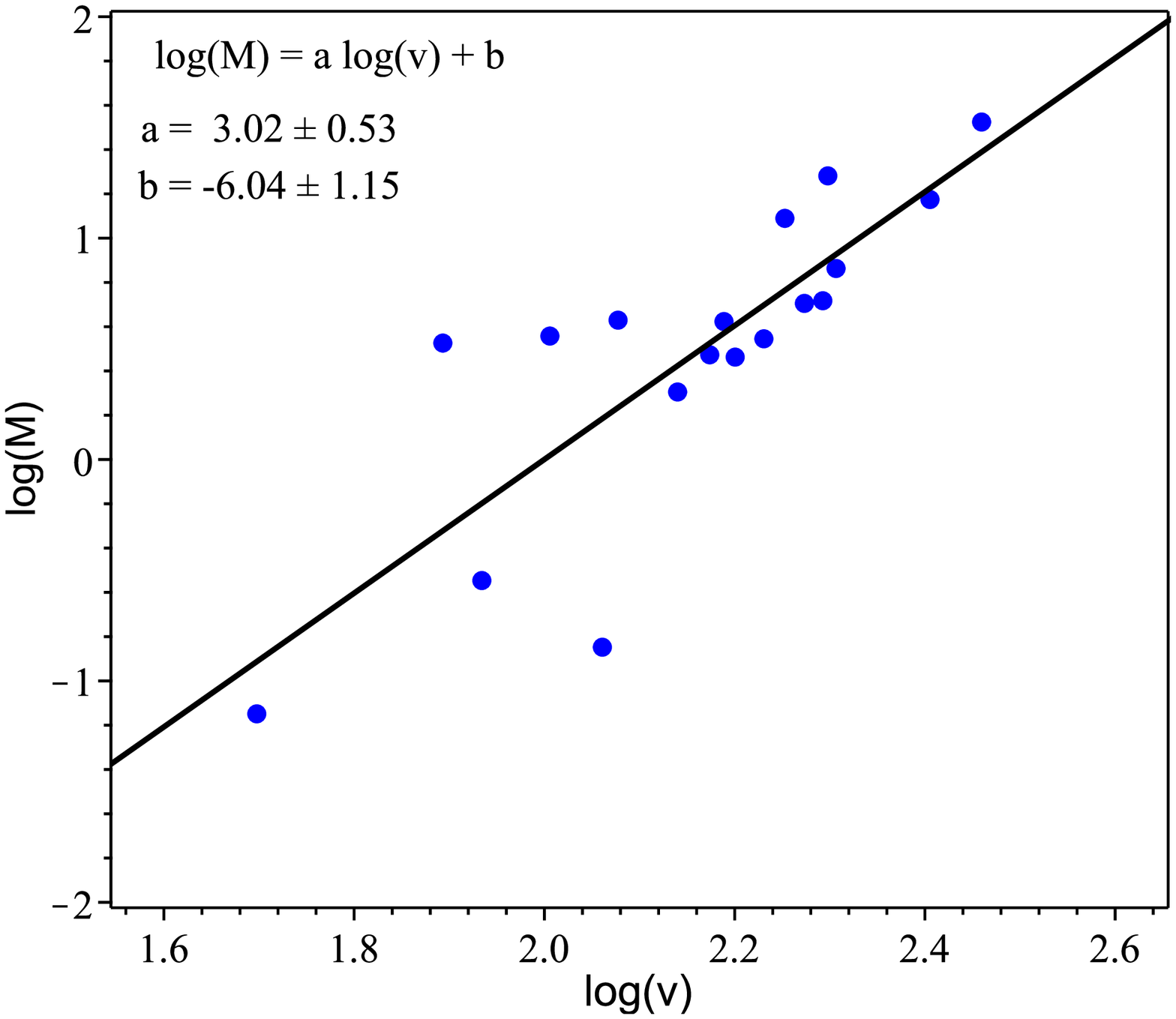}
\includegraphics[scale=0.4]{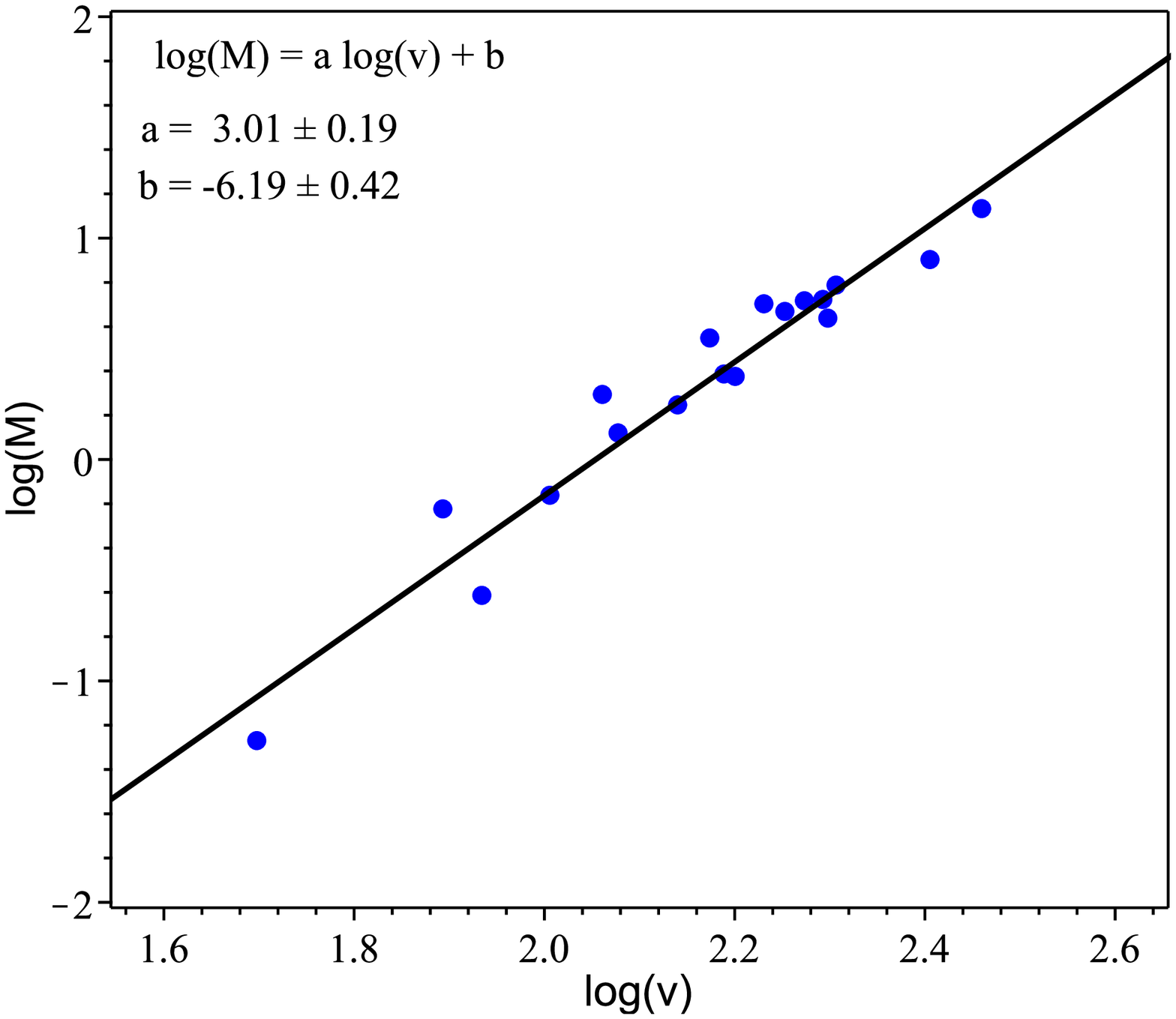}
\caption{(color online) The Tully-Fisher relation. {\it Left-top panel:} the observed B-band Tully-Fisher relation. Vertical
axis gives the base 10 logarithm of the observed luminosity (in units of $10^{10}\,L_{\odot}$, respectively $10^{10}\,M_{\odot}$) and the horizontal axis is
the base 10 logarithm of the last observed velocity (in km/s). {\it Left-bottom panel:} best fit Tully-Fisher relation parameterized by $\log(M)=a\log(v)+b$ in the
case of MOND. {\it Right-top panel:} Tully-Fisher best fit for the masses rezulted from the parametric model given by eq. (\ref{final}) using the mass
distribution (\ref{massnew}), respectively {\it Right-bottom panel:} Tully-Fisher relation obtained using the spherical version of the exponential disk
mass distribution (\ref{masssph}). The value of $M$ used in the plots is the total mass of a given galaxy: $M=M_*+M_{gas}$.}
\label{fig.5}
\end{figure*}

The empirical observational relation between the observed luminosity of a galaxy and the fourth power of the last observed velocity point is
known as the Tully-Fisher relation \cite{Tully}
\begin{equation}\label{TFrel}
L \propto v_{last}^4,
\end{equation}
which can be rewritten as
\begin{equation}\label{tully-fisher}
\log(M)=a\log(v)+b.
\end{equation}
In the figure \ref{fig.5} we have presented the observational Tully-Fisher relation (top-left panel) together with the fits of the parametric model given
by the equation (\ref{final}) using the mass distribution (\ref{massnew}) in the right-top panel and the spherical version of the exponential disk mass
distribution (\ref{masssph}) in the right-bottom panel, respectively. The left-bottom panel presents the Tully-Fisher relation coming from MOND mass predictions.

\section{Discussion and Conclusions}
In the presented paper we have considered the possible explanation of observed galactic rotation curves by the assumption that the observed effect of
the flatness can be explained by some alternative theory of gravity which introduces an extra term which we called $A(r)$.
This term can be treated as a deviation from the Newtonian limit of GR.

Our results are presented in the tables \ref{tab2} and \ref{tab3} together with the plots in the figures \ref{fig.2} and \ref{fig.4}.
Although we would like to think about
this contribution like something coming from a bit different geometry appearing in the modified Einstein field equations, it can be also thought as some extra
field, for example scalar one which recently has been considered as an agent of the cosmological inflation \cite{linde1, linde2, linde3}.
This choice for $A(r)$ in (\ref{func}) could be explained by considering two conformally related metrics
(the GR metric $g_{\mu\nu}$ and a "dark metric" $h_{\mu\nu}$) as proposed in \cite{sporea}. However, so far we have not been able to find a
suitable metric $h_{\mu\nu}$. It means, one needs to know a form of a lagrangian in the case of Palatini gravity in order to know the form of the dark metric.

Now on, we shall compare the new phenomenological model proposed in section \ref{intro2} for explaining flat galaxy rotation curves
with the widely accepted MOND model.

Let us start analyzing the predictions from the table \ref{tab2}. Comparing col. (7) and col. (3) from the table \ref{tab2} we observe that in all galaxies of the sample
(excepting NGC4826 and NGC7793) the predicted core radius $r_c$  is smaller than the galaxy length scale $R_0$. The same is true for MOND (excepting galaxies
NGC7793, DDO154 and IC2574). The ratio between the predicted MOND mass in col. (10) and the predicted mass in col. (6) is in the interval $(0.4, 8.1)$ such that for
13 out of 18 galaxies the MOND mass is higher.

The stellar mass-to-light ratio $M/L$ (denoted $\Upsilon_*$) is usually estimated in the literature \cite{McGaugh, Bell, Z09} by using color-to-mass-to-light ratio
relations (CMLR) of the type
\begin{equation}\label{CMLR}
\log\Upsilon_*^i=a_i+b_i\cdot color
\end{equation}
 $a,\,b$ are two parameters and $i$ is the band of the measured data. Then using the observed luminosity in the corresponding band, an
estimate of the stellar mass is
obtained. In \cite{McGaugh} the authors use CMLR and four stellar population synthesis models \cite{B03, P04, Z09, IP13} to compute
the stellar mass for a sample of 40 galaxies, including 13 of the THINGS galaxies used in this paper. Comparing our predicted stellar mass
from the table \ref{tab2}, col. (6) with the values from the table 3 in  \cite{McGaugh} and/or the values from the tables
3,4 in \cite{deBlok} we have
found that for 5 galaxies the predicted mass in col. (6) is in very good agreement, for 7 galaxies the mass is higher, while for 4 of the
galaxies the mass is slightly lower. Looking now at the values of col. (9) in the table \ref{tab2} and col. (5) in
the table \ref{tab3} we can say that the values of $\Upsilon_*$ are in agreement with what is expected based on stellar population
models \cite{McGaugh}. However, using the spherical mass distribution (\ref{masssph}) for LSB galaxies dose not result in good fits for
the rotational curves.

In col. (8) and col. (12) of the table \ref{tab2} the values of reduced $\chi^2$ are presented. These values were computed using
the standard definition: $\chi^2_r=\chi^2/(N-n)$, where $N$ is the number of observational velocity data points; $n$ is the number of parameters
to be fitted; and
\begin{equation}
\chi^2=\sum_i^N\left( \frac{V_i^{obs}-V_{model}(R_i)}{error_i} \right)^2.
\end{equation}

Taking all the above into account, one arrives to the conclusion that the new model (which does not assume the existence on any type of Dark Matter) proposed
in this paper gives very good flat rotation curves fits of the 18 THINGS galaxies in the data sample. Moreover, when compared with MOND the difference
between the two set of fits is small and thus one is not able to say which model is better than the other one for the explanation of the rotation curves.

We had not had any concrete theory in mind when we wanted to check our assumptions on the modification term $A(r)$. Since we have been influenced by the
results obtained by the others (briefly described in the section \ref{intro2}), we wanted to find much simpler modification apart MOND which also
provides a required shape of the galaxies curves. Therefore now, when
we have shown that observational data does not exclude the obtained result (\ref{vfinal}), it is stimulating to think about existing theories of gravity.

The proposed model presented in this paper (enclosed in eq. (\ref{vfinal})) can be viewed for now as a phenomenological model, until a concise theory of gravity
from which it can be derived, will be found or constructed. We started to tackle this task, thus working on a given theory of gravity which produces a simple
modification of the quadratic velocity is a topic of our current research.

\section*{Acknowledgements}

This work made use of THINGS, "The HI Nearby Galaxy Survey" (Walter et al. 2008).
We would like to thank Professors Fabian Walter and Erwin de Blok for helping us in obtaining the RC data from the THINGS catalogue.\\
AW is partially supported by the grant of the National Science Center (NCN) DEC- 2014/15/B/ST2/00089.
CS was partially supported by a grant of the Ministry of National Education and Scientific Research, RDI Programme for Space Technology and Advanced Research - STAR, project number 181/20.07.2017.

This article is based upon work from the COST Action CA15117, supported by COST (European Cooperation in Science and Technology). \\

\end{document}